\documentclass{article}
\usepackage[margin=1in]{geometry}
\usepackage{amsmath, amssymb, graphics, setspace}
\usepackage{tikz}

\usetikzlibrary{positioning}
\usepackage{hyperref}
\usepackage[style=ieee, citestyle=numeric-comp, backend=biber, url=false]{biblatex} 
\usepackage{multirow}
\usepackage{caption}
\usepackage{subcaption}
\usepackage[margin=1in]{geometry}
\usepackage{amsmath, amssymb, graphics, setspace}
\usepackage{tikz}

\usetikzlibrary{positioning}
\usepackage{hyperref}
\usepackage[style=ieee, citestyle=numeric-comp, backend=biber, url=false]{biblatex} 
\usepackage{multirow}
\usepackage{caption}
\usepackage{subcaption}
\usepackage{graphicx}

\newcommand{\mathsym}[1]{{}}
\newcommand{\unicode}[1]{{}}

\usepackage{boldline}

\usepackage{boldline}

\title{Reheating constraints on modified quadratic chaotic inflation}
\author{Sudhava Yadav$^1$, Rajesh Goswami$^2$, K.K Venkataratnam$^{1,\footnotemark{}}$ and Urjit A. Yajnik$^{3,4}$ 
\\$^1$ Department of Physics, Malaviya National Institute of Technology Jaipur, Jaipur 302017, India
\\$^2$ Department of Physics, National Institute of Technology Puducherry, Puducherry 609609, India
\\$^3$ Department of Physics, Indian Institute of Technology Bombay, Mumbai 400076, India
\\$^4$ Department of Physics, Indian Institute of Technology Gandhinagar, Gujarat 382424, India}

\begin{document}
\maketitle
\begin{abstract}
{The Reheating era of inflationary universe can be parameterized by  various parameters like reheating temperature \(T_{\text{re}}\), reheating duration \(N_{\text{re}}\) and average equation of state parameter \(\overline{\omega }_{\text{re}}\), which can be constrained by  observationally feasible values of  scalar power spectral amplitude \(A_{\text{s}}\) and spectral index \(n_{\text{s}}\). In this work, by considering the quadratic chaotic inflationary potential with logarithmic-correction in mass, we examine the reheating era in order to place some limits on model's parameter space. By investigating the reheating epoch using Planck 2018+BK18+BAO data, we show that even a small correction can make the quadratic chaotic model consistent with latest cosmological observations. We also find that the study of  reheating era helps to put much tighter constraints on model and effectively improves accuracy of model.}
\end{abstract}
\footnotetext{*Corresponding author}

\section{Introduction}\label{S1}
The inflationary paradigm
\cite{guth_inflationary_1981,starobinsky_new_1980,linde_new_1982,linde_chaotic_1983,riotto_inflation_2002}
is an exciting and influential epoch of the cosmological universe. It has come up as an aid to resolve a range of well-known cosmological problems like flatness, horizon and monopole problems of famous cosmological big bang theory. The semi-classical theory of inflation generates seeds for Cosmic Microwave Background anisotropy and Large Scale Structures in the late universe \cite{mukhanov_quantum_1981,starobinsky_dynamics_1982,guth_quantum_1985}. Inflation predicts adiabatic, gaussian and almost scale invariant density fluctuations, which are validated by CMB observations like Cosmic Background Explorer (COBE) \cite{smoot_structure_1992}, Wilkinson Microwave Anisotropy Probe (WMAP) \cite{dunkley_five-year_2009,komatsu_seven-year_2011} and Planck space probe \cite{ade_planck_2014,ade_planck_2014-1,ade_planck_2016,ade_planck_2016-1,aghanim_planck_2020,akrami_planck_2020-1}.

 In the realm of inflationary cosmology, a typical scenario involves the presence of a scalar field, which is referred to as the inflaton $(\phi )$, whose potential energy dominates the universe. In this picture, inflaton slowly rolls through its potential, and the coupling of quantum fluctuations of this scalar field with metric fluctuations is the source of primordial density perturbations called scalar perturbations. The tensor part of the metric has vacuum fluctuations resulting in primordial gravitational waves called tensor perturbations. During inflation, power spectra for both these perturbations depend on a potential called inflaton potential $V(\phi )$.
 
As Inflation ends, the universe reaches a highly nonthermal and frigid state with no matter content in it. However, the universe must be thermalized at extremely high temperature for big-bang nucleosynthesis (BBN) and baryogenesis. This is attained by {`}reheating{'}\cite{turner_coherent_1983,traschen_particle_1990,albrecht_reheating_1982,kofman_reheating_1994,kofman_towards_1997,drewes_kinematics_2013,allahverdi_reheating_2010}, transit between the inflationary phase and an era of radiation and matter dominance. 

There is no established science for reheating era and there is also a lack of direct observational data in favor of reheating. However, recent CMB data helped to obtain indirect bounds for various reheating parameters \cite{martin_observing_2015,martin_first_2010,dai_reheating_2014,martin_inflation_2006,adshead_inflation_2011,mielczarek_reheating_2011,cook_reheating_2015}, and those parameters are: the reheating temperature  \(\left(T_{\text{re}}\right)\), the effective equation of state (EoS) parameter during
reheating (\(\omega _{\text{re}}\)) and lastly, the reheating duration, which can be written in the form of number of e-folds \(\left(N_{\text{re}}\right.\)). It is challenging to bound the reheating temperature by LSS and CMB observations. However, its value is assumed to be higher than the electroweak scale for dark matter production at a weak scale. A lower limit has been set on reheat temperature i.e. \(\left(T_{\text{re}} \sim 10^{-2} GeV \right.\)) for a successful primodial nucleosynthesis (BBN) \cite{steigman_primordial_2007} and instantaneous reheating consideration allows us to put an upper bound i.e. \(\left(T_{\text{re}} \sim 10^{16} GeV \right.\)) for Planck's recent upper bound on tensor-to-scalar ratio (r). The value of second parameter, \(\omega _{\text{re}}\), shifts from \(-\frac{1}{3}\) to 1 in various scenarios. It is 0 for reheating generated by perturbative decay of a large inflaton and \(\frac{1}{3}\) for instantaneous reheating. The next parameter in line is the duration of reheating phase, \(N_{re}\). Generally, it is incorporated by giving a range of \(N_k\), the number of e-foldings from Hubble crossing of a Fourier mode  \(k\) to the termination of inflation. \(N_k\) has value in the range 46 to 70 in order to work out the horizon problem. These bounds arise by considering reheat temperature at electroweak scale and instantaneous reheating of the universe. A comprehensive analysis of higher bound on \(N_k\) is presented in \cite{Dodelson_2003,Liddle_2003}.

The relation between inflationary parameters and reheating can be derived by taking into consideration the progression of observable scales of cosmology from
the moment of their Hubble crossing during inflation to the current time. We can deduce relations among \(T_{\text{re}}, N_{\text{re}}\) and \(\omega
_{\text{re}}\), the scalar power spectrum amplitude \(\left(A_s\right.\)) and spectral index \(n_s\) for single-field inflationary models. Further, the constraints on \(T_{\text{re}}\) and \(N_{\text{re}}\) can be obtained from recent CMB data.

Although plenty of inflationary models have been studied in recent years\cite{martin_encyclopaedia_2013} and the inflationary predictions are in agreement with the recent CMB observations, there is still a need for a unique model. The most famous chaotic inflation with quadratic potential \(\left(\frac{1}{2}m^2\phi
^2\right)\) is eliminated by recent cosmological observations as it predicts large tensor perturbations due to large potential energy it has during inflaton at large field amplitudes. Hence, lowering the potential at higher field values can help getting rid of this obstacle. Numerous hypotheses in this vein have been put forth \cite{senoguz_chaotic_2008,enqvist_does_2014,ballesteros_radiative_2016,nakayama_polynomial_2013,nakayama_running_2010,pallis_kinetically_2015,kannike_dynamically_2015,boubekeur_current_2015,marzola_minimal_2016,racioppi_new_2018}. Radiative corrections provide an intriguing possibility \cite{senoguz_chaotic_2008,enqvist_does_2014,ballesteros_radiative_2016} where, generally, the quadratic potential gets flatter as result of running of inflaton's quartic coupling. This article will rather examine a straightforward scenario in which the mass exhibits a running behaviour described as \cite{kasuya_quadratic_2018}: 
\begin{equation} \label{1} 
    m^2(\phi )=m^2\left(1-\text{K} \ln  \left[\frac{\phi ^2}{M^2}\right]\right),
\end{equation}
where M is large mass scale and K is some positive constant. The positive K and the negative sign in above equation is a defining characteristic of dominance of the coupling of inflaton field to fermion fields. The reason for considering the
fermionic corrections over bosonic ones is that the earlier ones are likely to flatten the potential, lowering the value of tensor to scalar ratio(r). Hence, making the potential compatible with recent observations\cite{ade2021improved}. In contrast, the bosonic ones steepen the potential and increase the r value\cite{enqvist_does_2014,ballesteros_radiative_2016,ahmed2016quantum}. Another interesting way to make such models compatible with observation is by extension of standard model as done in Ref. \cite{Borah:2019bdi,Ghoshal:2022zwu}.\\
Reheating is well known technique of constraining the inflationary models. There are various ways to analyse the reheating phase as available in literature e.g. one stage reheating study \cite{goswami_reconciling_2018,cook_reheating_2015}, two stage reheating study \cite{goswami_reheating_2020,Maity:2019ltu}. In Ref. \cite{Drees:2021wgd} reheating was analysed through perturvative decay of inflaton to either bosonic or fermionic states through trilinear coupling. Considering one stage reheating technique of constraining the models, we use various reheating parameters to put much tighter bounds
on parameter space of quadratic chaotic inflationary model with a logarithmic-correction in mass in light of Planck 2018+BK18+BAO data \cite{aghanim_planck_2020,akrami_planck_2020-1,ade2021improved}. By demanding \(T_{\text{re}}> 100\) GeV for production of weak-scale dark matter and working in plausible range of average equation of state (EoS) parameter ($-\frac{1}{3} \le \overline{\omega }_{\text{re}} \le 1$), we employ the derived relation between inflationary and reheating parameters and observationally feasible values of \(A_s\), \(n_s\) and r to place a limit on model's parameter space. It is a helpful and fairly new tool for putting relatively tighter constraints on the model and reducing its viable parameter space, providing significant improvement in accuracy of the model. Additionally, this technique well differentiate various inflation models as they can have the same forecasts for \(n_s\) and r, but definitely not for the same \(\omega _{\text{re}}\), as the tightened constraints
on \(n_s\) will result in an increasingly narrow permitted range of \(\omega _{\text{re}}\) for a particular inflationary model.\\
The organization of this paper is as follows: In Sec. \ref{S2} we discuss the dynamics and predictions of slow-roll inflation. We also derived the expressions for \(T_{\text{re}}\)
and \(N_{\text{re}}\) as a function of \(\overline{\omega }_{\text{re}}\) and other inflationary parameters like \(\text{$\Delta $N}_k\) and \(V_{\text{end}}\). In section \ref{S3}, the Subsec. \ref{S3.1} has our recreated data for reheating scenario of simple quadratic chaotic potential. In Subsec. \ref{S3.2}, we discussed the various field domains within which inflation can occur for quadratic chaotic potential with logarithmic correction in mass and then we parameterized reheating for this model using \(T_{\text{re}}\) and \(N_{\text{re}}\) as
a function of the scalar spectral index \(n_s\) for different \(\omega _{\text{re}}\). We have also examined the observational limits and reheating parameters for both these models using Planck 2018+BK18+BAO data in Sec. \ref{S3}. Sec. \ref{S4} is reserved for discussion and conclusions.\\
 We will be working with $\hbar$ \(=c=1\) units and the values of some standard parameters used are  reduced Planck{'}s mass \(M_{P }=\sqrt{\frac{1}{8\pi  G}}\) \(=\text{}\)2.435 $\times
$ \(10^{18}\) GeV, the redshift of matter radiation equality \(z_{\text{eq}} \approx\) 3400, \(g_{\text{re}}\) $\approx $ 100 \cite{dai_reheating_2014} and the present value of Hubble parameter $H_o = 100$h km $s^{-1 }$ $\text{Mpc}^{-1}$
 with h = 0.68 \cite{aghanim_planck_2020,akrami_planck_2020-1}

\section{Parameterizing reheating in slow-roll inflationary models}\label{S2}
Reheating phase can be parameterized by assuming it been dominated by some fluid \cite{martin_inflation_2003}
of energy density \(\rho\) with pressure P and equation of state(EoS) parameter \(\omega _{\text{re}}=\frac{P}{\rho }\) where
\begin{equation} \label{2}
\rho =\frac{\dot{\phi }^2}{2}+V(\phi ),\;\;\;\;\;\;\;\;\;   P =\frac{\dot{\phi }^2}{2}-V(\phi ).
\end{equation}
The continuity equation gives
\begin{equation} \label{3}
\dot{\rho }+3H(P+\rho)=0,
\end{equation}

\begin{equation} \label{4}
\dot{\rho }+3H\rho \left(\omega _{\text{re}}+1\right)=0.
\end{equation}
We analyze the dynamics of inflation by considering inflaton \(\phi\) with potential \(V(\phi )\)  evolving slowly with slow-roll parameters \(\epsilon\) and \(\eta\). The approximation of Friedman equation using slow-roll conditions give
\begin{equation}\label{5}
3H\dot{\phi }+V'(\phi)=0,
\end{equation}
\begin{equation} \label{6}
H^2=\frac{V(\phi )}{3M_P^2},
\end{equation}
where prime \((')\) denotes derivative w.r.t \(\phi\) and H = \(\frac{\dot{a}}{a}\) is Hubble parameter. The definition of slow-roll parameter give
\begin{equation} \label{7}
\epsilon =\frac{M_P^2}{2}\left(\frac{V'}{V}\right)^2,\;\;\;\;\;\;\;\;\;\;\;   \eta =M_P^2\left(\frac{V''}{V}\right).
\end{equation}
The scalar spectral index \(n_s\), tensor spectral index \(n_T\) and tensor to scalar ratio \(r\) in terms of above slow-roll parameters satisfy the relations
 \begin{equation} \label{8}
 \text{   }n_s=1-6\epsilon +2\eta,\;\;\;\;\;\;\;n_T=-2\epsilon,\;\;\;\;\;\;\;r=16\epsilon.
\end{equation}
Now, the number of e-foldings in between Hubble crossing of mode \(k\) and termination of inflation denoted by subscript ``end" can be given as
\begin{equation} \label{9}
\Delta N_k=\ln \left(\frac{a_{\text{end}}}{a_{\text{k}}}\right)= \frac{1}{M_P^2}\int_{\phi _{\text{end}}}^{\phi _k} \frac{V}{V'} \, d\phi,
\end{equation}
where \(a_{k}\) and \(\phi _{k }\) represents value of scale factor and inflaton at the point of time when \(k\) crosses the Hubble radius. The later part of eq. (\ref{9}) is obtained using the slow-roll approximations \(\ddot{\phi } \ll 3H\dot{\phi }\) and \(V(\phi ) \gg \dot{\phi }^2\). Similarly,
\begin{equation} \label{10}
N_{\text{re}}=\ln \left(\frac{a_{\text{re}}}{a_{\text{end}}}\right), 
\end{equation}
Here the quantity \(N_{\text{re}}\) encrypts both, an era of preheating \cite{kofman_towards_1997,boyanovsky_preheating_1996,kofman_reheating_1998,felder_instant_1998,giudice_cosmological_2001,desroche_preheating_2005} 
as well as later thermalization process. An energy density controls the Universe's subsequent evolution and can be written as
\begin{equation} \label{11}
\text{    }\rho _{\text{re}}=\frac{\pi ^2}{30}g_{\text{re}}T_{\text{re}}^4,
\end{equation}
where \(g_{\text{re}}\) gives the actual count of relativistic species at termination of reheating epoch and \(T_{\text{re}}\) is the reheating temperature. Now, in view of eq. (\ref{3})

\begin{equation} \label{12}
\rho _{\text{re}}=\rho _{\text{end}}e^{-3N_{\text{re}}\left(1+\overline{\omega }_{\text{re}}\right)},
\end{equation}
\begin{equation} \label{13}
\text{where }{}{} \overline{\omega }_{\text{re}} =<\omega > =\frac{1}{N_{\text{re}}}\int _{N_{\text{end}}}^N\omega _{\text{re}}(N)dN
\end{equation}
Here \(\overline{\omega }_{\text{re}}\) is average EoS parameter during reheating \cite{martin_observing_2015,goswami_reconciling_2018}. Now, eq. (\ref{12}) can be recast as

\begin{equation} \label{14}
\frac{a_{\text{re}}}{a_{\text{end}}}=e^{N_{\text{re}}}=\left(\frac{\rho _{\text{re}}}{\rho _{\text{end}}}\right)^{-\frac{1}{3(1+\overline{\omega
}_{\text{re}})}}.
\end{equation}
Using eq. (\ref{11}) and eq. (\ref{14}), reheating e-folds \(N_{\text{re}}\) can be written as

\begin{equation} \label{15}
N_{\text{re}}=\frac{1}{3\left(1+\overline{\omega }_{\text{re}}\right)}\left\{\ln \left(\frac{3}{2}V_{\text{end}}\right)-\ln \left(\frac{\pi ^2}{30}g_{\text{re}}\right)\right\}-\frac{4}{3\left(1+\overline{\omega
}_{\text{re}}\right)}\ln \left(T_{\text{re}}\right).
\end{equation}
For some physical scale \(k\), the observed wavenumber `\(\frac{k}{a}\)' can be given in terms of above known quantities and the redshift during matter-radiation equality epoch (\(z_{\text{eq}}\)) as \cite{goswami_reconciling_2018}

\begin{equation} \label{16}
H_k=\frac{k}{a_k}=\left(1+z_{\text{eq}}\right)\frac{k}{a_o}\rho _{\text{re}}^{\frac{3~\overline{\omega }_{\text{re}}-1}{12\left(1+\overline{\omega
}_{\text{re}}\right)}}\rho _{\text{eq}}{}^{-\frac{1}{4}}\left(\frac{3}{2}V_{\text{end}}\right)^{\frac{1}{3\left(1+\overline{\omega }_{\text{re}}\right)}}e^{\Delta
N_k}.
\end{equation}
Using eq. (\ref{16}), \(\Delta N_k\) can be given as
\begin{equation} \label{17}
\Delta N_k=\ln  H_k-\ln \left(1+z_{\text{eq}}\right)-\ln \left( \frac{k}{a_o}\right)-\frac{1}{\left.3(\overline{\omega
}_{\text{re}}+1\right)}\ln \left(\frac{3}{2}V_{\text{end}}\right)-\frac{3~\overline{\omega }_{\text{re}}-1}{3\left(1+\overline{\omega
}_{\text{re}}\right)}\ln \left(\rho _{\text{re}}^{\frac{1}{4}}\right)+\ln \left(\rho _{\text{eq}}^{\frac{1}{4}}\right).
\end{equation}
Inverting eq. (\ref{17}), and using eq. (\ref{11}) one can get a mutual relation among the numerous parameters introduced,

\begin{equation} \label{18}
\ln  (T_{\text{re}})=\frac{3~\left(1+\overline{\omega }_{\text{re}}\right)}{3~ \overline{\omega }_{\text{re}}-1}\left\{\ln  H_k-\ln
\left(1+z_{\text{eq}}\right)-\ln  \frac{k}{a_o}-\Delta N_k+\ln \left(\rho _{\text{eq}}^{\frac{1}{4}}\right)\right\}
-\frac{1}{3~\overline{\omega }_{re}-1}\ln \left(\frac{3}{2}V_{end}\right)-\frac{1}{4}\ln \left(\frac{\pi ^2}{30}g_{re}\right).
\end{equation}
The expression for \(T_{\text{re}}\) from eq. (\ref{15})  is substituted in eq. (\ref{18}) to get the expression for \(N_{\text{re}}\) as mentioned below
\begin{equation} \label{19}
N_{\text{re}}=\frac{1}{3~\overline{\omega }_{\text{re}}-1}\ln \left(\frac{3}{2}V_{\text{end}}\right)+\frac{4}{3~\overline{\omega }_{\text{re}}-1}\left\{\ln
\left(\frac{k}{a_o}\right)+\Delta N_k+\ln \left(1+z_{\text{eq}}\right)-\ln \left(\rho _{\text{eq}}^{\frac{1}{4}}\right) -\ln  H_k \right\}
\end{equation}
eq. (\ref{18}) and eq. (\ref{19}) are the two key relationships for parameterizing reheating in slow-roll inflationary models.
\section{Inflationary models}\label{S3}
\subsection{Quadratic Chaotic inflationary model}\label{S3.1}

We are first considering simple quadratic chaotic potential before moving to its modified form. 
The quadratic chaotic potential \cite{linde_chaotic_1983} has the form

\begin{equation} \label{20}
V=\frac{1}{2}m^2\phi ^2.
\end{equation}
The reheating study of this potential was already done in \cite{goswami_reconciling_2018} in light of Planck's 2015 data. We are recreating the data by doing the similar study using Planck 2018+BK18+BAO data. 
Using  eq. (\ref{7}) slow-roll parameters for this potential can be given as\\
\begin{equation} \label{21}
\epsilon =\eta =\frac{2 M_P^2}{\phi ^2}.
\end{equation}
The Hubble parameter during the crossing of Hubble radius by scale \(k\) for this model can be written as

\begin{equation} \label{22}
H_k^2=\frac{1}{M_P^2}\left(\frac{V_k}{3-\epsilon _k}\right)=\frac{1}{2M_P^2}\left(\frac{m^2\phi _k^2}{3-2 \frac{M_P^2}{\phi ^2}}\right).
\end{equation}
where \(\phi _k\), \(\epsilon _k\) and \(V_k\) respectively represent the inflaton field, slow-roll parameter and potential during crossing of Hubble radius by mode \(k\).\\
Using the condition \(\epsilon =1\), defining end of inflation, in eq. (\ref{21}), we obtained \(\frac{\phi _{\text{end}}^2}{M_P^2}=2\)\\
Now, corresponding to pivot scale \(k_*\), used in Planck collaboration, \(\frac{k_*}{a_o}=0.05 Mpc^{-1}\), consider the mode \(k_*\) crossing the hubble radius at a point  where the field has achieved the value \(\phi _*\)  during inflation. The remaining number of e-folds persist subsequent to crossing of hubble radius by \(k_*\)  are
\begin{equation} \label{23}
\Delta N_*\simeq \frac{1}{M_P^2}\int_{\phi _{\text{end}}}^{\phi _*} \frac{V}{V'} \, d\phi \\
\\
=\left[\left(\frac{\phi _*}{2M_P}\right)^2-\frac{1}{2}\right].
\end{equation}
The spectral index for this model can be easily obtained using eq. (\ref{8}) as

\begin{equation} \label{24}
n_s=1-8\left(\frac{M_P^2}{\phi _*^2}\right).
\end{equation}
Now, the formulation for tensor-to-scalar ratio from eq. (\ref{8}) gives

\begin{equation} \label{25}
r=32\frac{ M_P^2}{\phi _*^2}.
\end{equation}
Moreover, this model yields the relation
\begin{equation} \label{26}
H_*=\pi  M_P\sqrt{16A_s\frac{ M_P^2}{\phi _*^2}}.
\end{equation}
The relation of field \(\phi\) and \(H\) eq. (\ref{6}), and the condition for termination of inflation as used in eq. (\ref{23}), along with eq. (\ref{26}) gives expression for \(V_{\text{end}}\) as
\begin{equation} \label{27}
V_{\text{end}}(\phi )=\frac{1}{2}m^2\phi _{\text{end}}^2=\frac{3H_*^2M_P^2\phi _{\text{end}}^2}{\phi _*^2}=\frac{6H_*^2M_P^4}{\phi _*^2}.
\end{equation}
\begin{figure}[!h]
    \centering
   {\includegraphics[width=\textwidth]{ 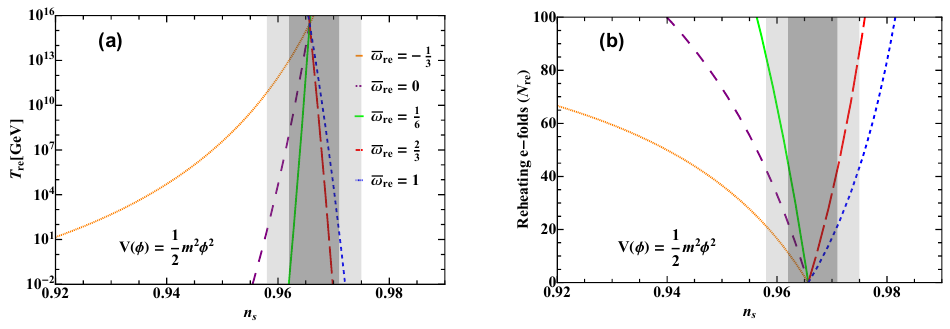}}
    \caption{The plots for (a) \(T_{\text{re}}\) and (b) \(N_{\text{re}}\) versus \(n_{s }\) for the quadratic chaotic model
\(V(\phi )=\frac{1}{2}m^2\phi ^2\) for different values of { }\(\overline{\omega }_{\text{re}}\) : \(\overline{\omega }_{\text{re}}\)=\(-\frac{1}{3}\)(dotted orange), \(\overline{\omega }_{\text{re}}\)= 0(medium dashed purple), \(\overline{\omega
}_{\text{re}}\)=\(\frac{1}{6}\)(solid green), \(\overline{\omega }_{\text{re}}\)=\(\frac{2}{3}\)(large dashed red), \(\overline{\omega }_{\text{re}}\)= 1(small dashed blue). The regions with light and dark gray shades, respectively, represent the \(\text{2$\sigma $}\) and \(\text{1$\sigma $}\) bounds on \(n_s\) from Planck 2018 (TE,EE,TT+Low E+Lensing)+BK18+BAO data \cite{aghanim_planck_2020,akrami_planck_2020-1,ade2021improved}}.
    \label{F2}
\end{figure}

Now, the expressions for  $\Delta $\(N_*\), \(r\), \(H_*\) and \(V_{\text{end} }\) as a function of \(n_{s\text{  }}\) can be obtained by putting the value of \(\phi_*\) from eq. (\ref{24}) in  eqs. (\ref{23}), (\ref{25}), (\ref{26}) and (\ref{27}), and then these expressions along with eqs. (\ref{18}) and (\ref{19}) gives number of reheating e-folds \(N_{\text{re}}\) and reheating temperature \(T_{\text{re}}\). Planck's 2018 value of \(\text{
 }A_{s }=2.1\times 10^{-9}\) and computed value of \(\rho
_{\text{eq}}^{\frac{1}{4}} = 10^{-9}\)GeV \cite{aghanim_planck_2020,akrami_planck_2020-1} have been used for calculation. The \(N_{\text{re}}\) and \(T_{\text{re}}\) versus \(n_s\) plots, along with Planck-2018 \(1\sigma\) bound on  \(n_s\) i.e. $(0.962 \le  n_s \le  0.971)$ (dark gray) and \(2\sigma\) bound on  \(n_s\) i.e. $(0.958 \le  n_s \le  0.975)$ (light gray), for this model are presented graphically in figure \ref{F2} for a range of average EoS parameter during reheating.\\
By demanding \(T_{\text{re}} \ge 100\) GeV for production of weak-scale dark matter and solving eqs. (\ref{18}) and (\ref{24}), the bounds on \(n_s\) are obtained and are reflected on eq. (\ref{23}) and eq. (\ref{25}) to obtain bounds on  $\Delta $\(N_*\) and r. All the obtained bounds are shown in table \ref{T1}. For this model the bounds on \(n_s\) lies inside Planck 2018+BK18+BAO \(2\sigma\) bound demanding \(\overline{\omega }_{\text{re}}\) lies
in the range ($0 \le \overline{\omega }_{\text{re}} \le 1$) and the corresponding range for r is (0.169 $\ge $ r $\ge $ 0.117) while if we demand \(n_{s }\) to lie within \(1\sigma\) bound then the allowed range of
\(\overline{\omega }_{\text{re}}\) is  ($0.127 \le \overline{\omega }_{\text{re}} \le 1$) and the corresponding r values are (0.152 $\ge $ r $\ge $ 0.117). Within these ranges of \(\overline{\omega }_{\text{re}}\) the tensor-to-scalar
ratio (r) is greater than the Planck 2018 and BK18 bound (\(r<0.036)\) \cite{ade2021improved}.\\
\begin{table}[h]
\caption{The permissible range for values of $n_s$, $\Delta N_*$ and $r$ for Quadratic Chaotic inflationary potential $\left(\frac{1}{2}m^2\phi
^2\right)$ by demanding $T_{re} \geq 100GeV$}\label{T1}%
    \centering
    \begin{tabular}{|c|c|c|c|c|c|}
        \hline
          \multirow{5}{*}{$\left(\frac{1}{2}m^2\phi
^2\right)$} & Average Equation of state & $n_s$ & $\Delta N_*$ & $r$ \\
        \cline{2-5}
        & $-\frac{1}{3} \le \overline{\omega }_{re} \le 0$ & $0.926 \le n_s \le 0.958$ & $26.47 \le \Delta N_* \le 47.45$ & $0.297  \ge r \ge 0.166$ \\
         \cline{2-5}
        & $0 \le \overline{\omega }_{re} \le \frac{1}{6}$ & $0.958 \le n_s \le 0.963$ & $47.45 \le \Delta N_* \le 53.38$ & $0.166  \ge r \ge 0.148$ \\
         \cline{2-5}
        & $\frac{1}{6} \le \overline{\omega }_{re} \le \frac{2}{3}$ & $0.963 \le n_s \le 0.969$ & $53.38 \le \Delta N_* \le 63.99$ & $0.148 \ge r \ge 0.124$ \\
        \cline{2-5}
        & $\frac{2}{3} \le \overline{\omega }_{re} \le 1$ & $0.969 \le n_s \le 0.971$ & $63.99 \le \Delta N_* \le 68.10$ & $0.124  \ge r \ge 0.117$ \\
        \hline   
    \end{tabular}  
\end{table}
From table \ref{T1}, we can see that all the r values for this model are greater than the combined Planck 2018+BK18+BAO bound (\(r<0.036)\) \cite{ade2021improved}. Hence, this model is incompatible with the data for any choice of $\overline{\omega }_{re}$ taken.
\subsection{Modified quadratic chaotic inflation }\label{S3.2}
 The quadratic chaotic inflationary potential with logarithmic-correction in mass term has the form
\cite{ballesteros_radiative_2016,kasuya_quadratic_2018}

\begin{equation} \label{28}
V(\phi )=\frac{1}{2}m^2\left(1-\text{K} \ln\frac{\phi ^2}{M_P^2}\right)\phi ^2=(M')^4\left(1-\text{K} \ln\frac{\phi ^2}{M_P^2}\right)\frac{\phi ^2}{M_P^2},
\end{equation}
where $(M')^4=m^2M_P^2/2$ and K is some positive constant. The positive K is a defining characteristic of dominance of fermion couplings. This work is inspired by Ref. \cite{kasuya_quadratic_2018,senoguz_chaotic_2008}, where the inflationary scenario of this potential was studied. We are considering this potential in context of reheating in light of Planck 2018+BK18+BAO data.
\begin{figure}[!h]
 \centering
    \includegraphics[width=0.6\textwidth]{ 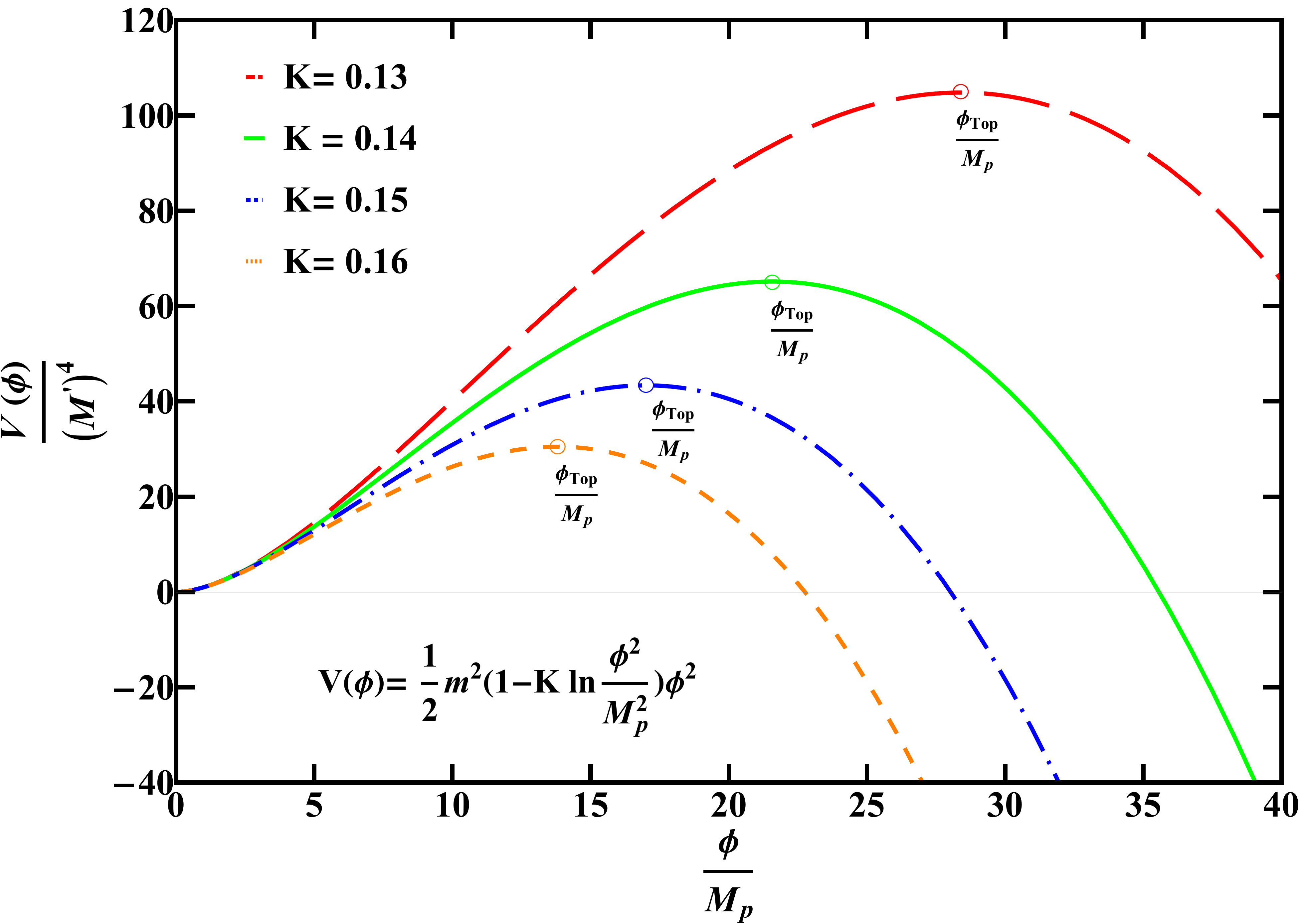}
    \caption{The plot of potential versus $\frac{\phi}{M_P}$ for quadratic
chaotic model with corrected mass \(V(\phi )=\frac{1}{2}m^2\left(1-\text{K} \ln \frac{\phi ^2}{M_P^2}\right)\phi ^2\) for different values
of {K} : {K}=0.13 (large dashed red), {K}=0.14 (solid green), {K}=0.15 (dot-dashed blue), {K}=0.16 (small dashed orange). The small circle on each curve signifies the maxima of potential in each case and the corresponding field value at maxima is $\frac{\phi_{Top}}{M_P}$.} 
\label{F6} 
\end{figure}\\
We will start our discussion with various field domains \cite{martin_encyclopaedia_2013} within which inflationary phenomena may occur for above potential. It is evident that the above-mentioned potential eq. (\ref{28}) does not exhibit positive definiteness for all values of the field (${\phi}$). The value of this potential becomes negative after a specific point
\begin{equation} \label{42}
\frac{\phi_{V=0}}{M_P}=\sqrt{e^\frac{1}{\text{K}}}.
\end{equation}
The model can only be defined within a specific regime i.e., $\phi < \phi_{V=0}$. On the contrary, the highest point of the potential function, where \(V'=0\) (or can say $\epsilon=0)$, corresponds to field value given as:
\begin{equation} \label{43}
\frac{\phi_{V'=0}}{M_P}=\frac{\phi_{Top}}{M_P}=\sqrt{e^\frac{1-\text{K}}{\text{K}}},
\end{equation}
The model has a sense provided the correction term doesn't have its dominance on the potential, hence the suitable regime is  $\phi <\phi_{Top}< \phi_{V=0}$. We have ignored the additional stabilizing terms coming from stability of potential at higher field region as inflation is taking place for field region below the maxima($\phi_{Top}$), which is a stable potential region. Most naturally, inflation begins at an energy density closer to Planck scale and it's observable part takes place at much lower energy density. After the initial inflationary phase if there exist a field region close to maxima, eternal inflation happens and that region will always be dominating. See e.g. refs \cite{senoguz_chaotic_2008,linde2005particle,vilenkin1983birth} for more discussion on this point.
The potential versus $\frac{\phi}{M_P}$ plot for four different values of K is depicted in figure \ref{F6}. From figure \ref{F6} it can be seen that each K has specific viable regime in which the model is defined and have a sense and we will be working in these regions only.\\
Now moving further, the slow-roll parameters for this potential can be given as
\begin{equation} \label{29}
\epsilon =2M_P^2\left(\frac{ 1-\text{K}\left(1+ \ln \frac{\phi ^2}{M_P^2}\right)}{\phi (1-\text{K} \ln \frac{\phi ^2}{M_P^2}}\right)^2,
\end{equation}

\begin{equation} \label{30}
\eta =\frac{2M_P^2 \left(-3 \text{K}+1-\text{K} \ln \frac{\phi ^2}{M_P^2}\right)}{\phi ^2 \left(1-\text{K} \ln \frac{\phi
^2}{M_P^2}\right)}.
\end{equation}
After substituting the values of $\epsilon $ eq. (\ref{29}) and $\eta $ eq. (\ref{30}) in eq. (\ref{8}), we can write scalar spectral index as

\begin{equation}\label{31}
n_s=\frac{-4 \left(2-3\text{K}+3 \text{K}^2\right) +\frac{\phi ^2}{M_P^2}-2\text{K}\left(-8+6 \text{K} +\frac{\phi ^2}{M_P^2}\right) \ln \frac{\phi
^2}{M_P^2}+\text{K}^2 \left(-8 +\frac{\phi ^2}{M_P^2}\right) \ln \left[\frac{\phi ^2}{M_P^2}\right]^2}{\frac{\phi ^2}{M_P^2} \left(-1+\text{K}\ln
\frac{\phi ^2}{M_P^2}\right)^2}.
\end{equation}
The Hubble parameter during the crossing of Hubble radius by scale \(k\) can be written as
\begin{equation} \label{32}
H_k^2=\frac{1}{M_P^2}\left(\frac{V_k}{3-\epsilon _k}\right)=\left(\frac{\frac{1}{2}m^2\left(1-\text{K} \ln \frac{\phi _k^2}{M_P^2}\right)\frac{\phi
_k^2}{M_P^2}}{3-2M_P^2\left(\frac{ 1-\text{K}\left(1+ \ln \frac{\phi ^2}{M_P^2}\right)}{\phi \left(1-\text{K} \ln \frac{\phi ^2}{M_P^2}\right)}\right)^2}\right).
\end{equation}
Using the condition \(\epsilon =1\) defining end of inflation, we have obtained \(\frac{\phi _{\text{end}}}{M_P}\) for different values of K.  The remaining number of e-folds persist subsequent to crossing of Hubble radius by \(k_*\) till the termination of inflationary epoch can be given as

\begin{equation} \label{33}
\text{$\Delta $}N_*\simeq \frac{1}{M_P^2}\int_{\phi _{\text{end}}}^{\phi _*} \frac{V}{V'} \, d\phi _*=\frac{1}{2M_P^2}\int_{\phi _{\text{end}}}^{\phi _*} \frac{ \left(1-\text{K {ln}} \frac{\phi ^2}{M_P^2}\right)\phi}{
1- \text{K}\left(1+\text{ln}\frac{\phi ^2}{M_P^2}\right)} \, d\phi.
\end{equation}
Defining \(\frac{\phi _*}{M_P}=x\). The spectral index \(n_s\) eq. (\ref{31}), at \(\phi =\phi _*\) in terms of $x$ will have the form

\begin{equation} \label{34}
n_s=\frac{-4 \left(2-3\text{K}+3 \text{K}^2\right) +x^2-2\text{K}\left(-8+6\text{K} +x^2\right) \ln x^2+\text{K}^2 \left(-8 +x^2\right)
\ln \left[x^2\right]^2}{x^2 \left(-1+\text{K}\ln x^2\right)^2}.
\end{equation}
\begin{figure}[!t]
    \centering
    {\includegraphics[width=\textwidth]{ 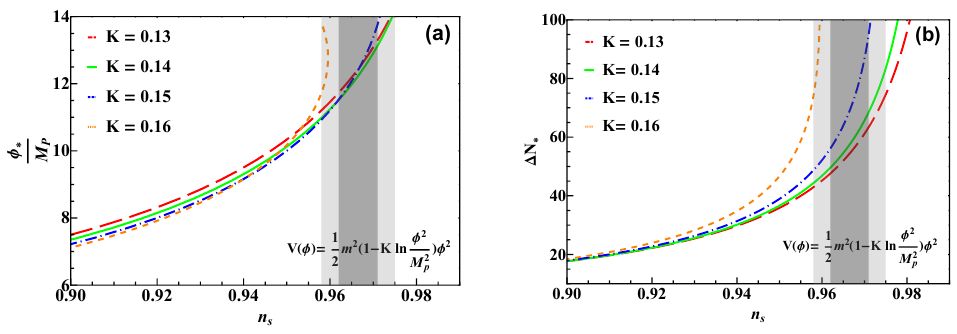}}
    \caption{The plots of (a) \(\frac{\phi _*}{M_P}\) and (b) \(\text{$\Delta $N}_*\) versus \(n_{s }\) for quadratic
chaotic model with corrected mass \(V(\phi )=\frac{1}{2}m^2\left(1-\text{K} \ln \frac{\phi ^2}{M_P^2}\right)\phi^2\) for different values
of {K}. The shaded regions and color codings are same as in figure \ref{F2} and \ref{F6} respectively.}
\label{F3}
\end{figure}

\begin{figure}
    \centering

\includegraphics[width=\textwidth]{ 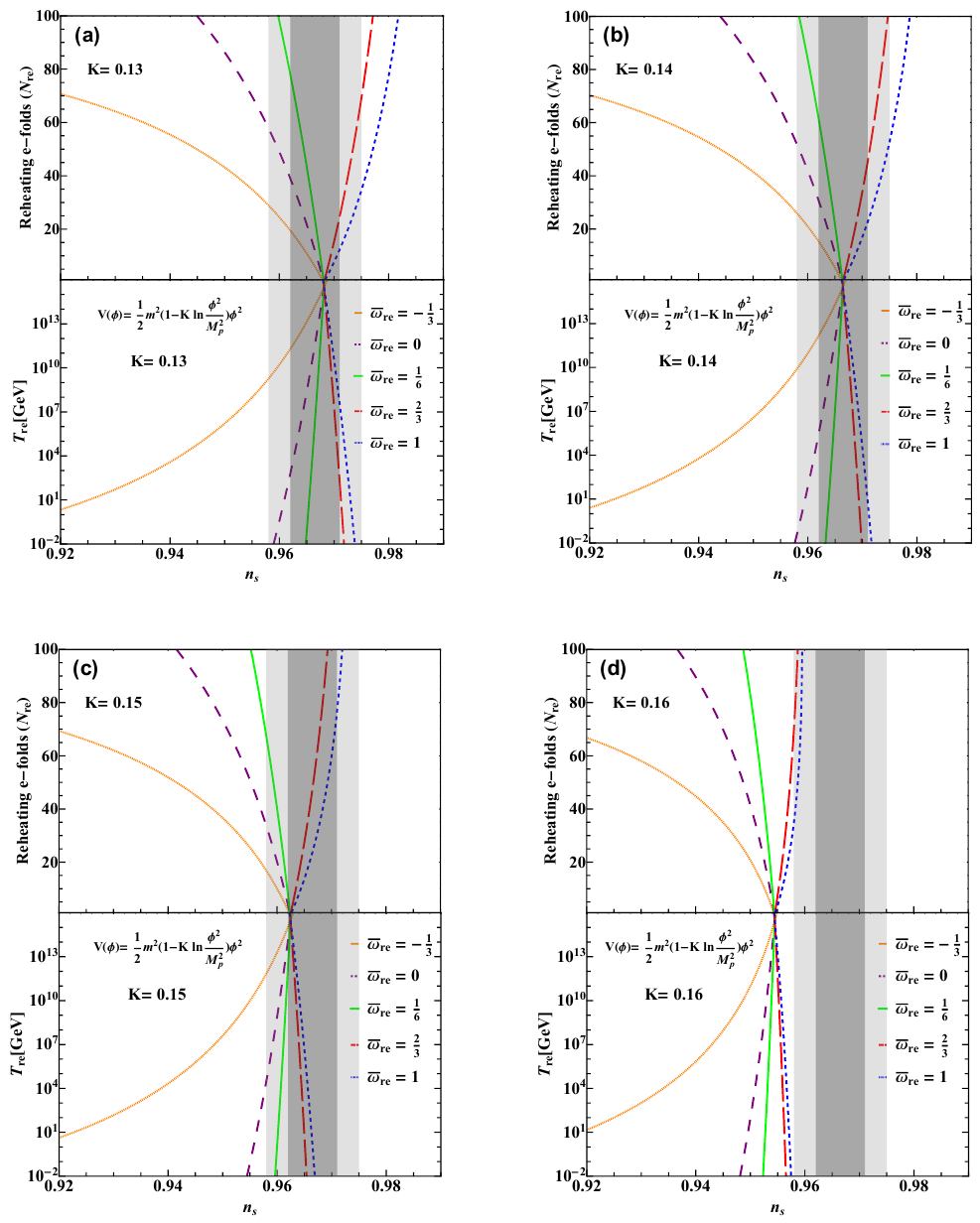}
    \caption{The plots for \(T_{\text{re}}\) and \(N_{\text{re}}\) versus spectral index (\(n_{s }\)) for quadratic chaotic
model with corrected mass \(V(\phi )=\frac{1}{2}m^2\left(1-\text{K} \ln \frac{\phi ^2}{M_P^2}\right)\phi ^2\) for different {K} and \(\overline{\omega }_{\text{re}}\) values. The shaded regions and color codings are same as in figure \ref{F2}}
    \label{F4}
\end{figure}

The variation of \(\frac{\phi _*}{M_P}\) and \(\text{$\Delta $N}_*\) with \(n_{s }\) using  eq. (\ref{34}) and  eq. (\ref{33}) for 4 different values of K are shown in figure \ref{F3}a and \ref{F3}b respectively. Further in this model, we can write the tensor - to - scalar ratio and $H_*$ as

\begin{equation} \label{35}
r=32\left(\frac{1-\text{K}\left(1+ \ln x^2\right) }{x\left(1-\text{K} \ln x^2\right) }\right)^2.
\end{equation}

\begin{equation} \label{36}
H_*=4\pi  M_P\sqrt{A_s}\left(\frac{1-\text{K}\left(1+ \ln x^2\right)}{x\left(1-\text{K} \ln x^2\right)}\right).
\end{equation}

\begin{figure}[!t]
    \centering
     \includegraphics[width=0.75\textwidth]{ 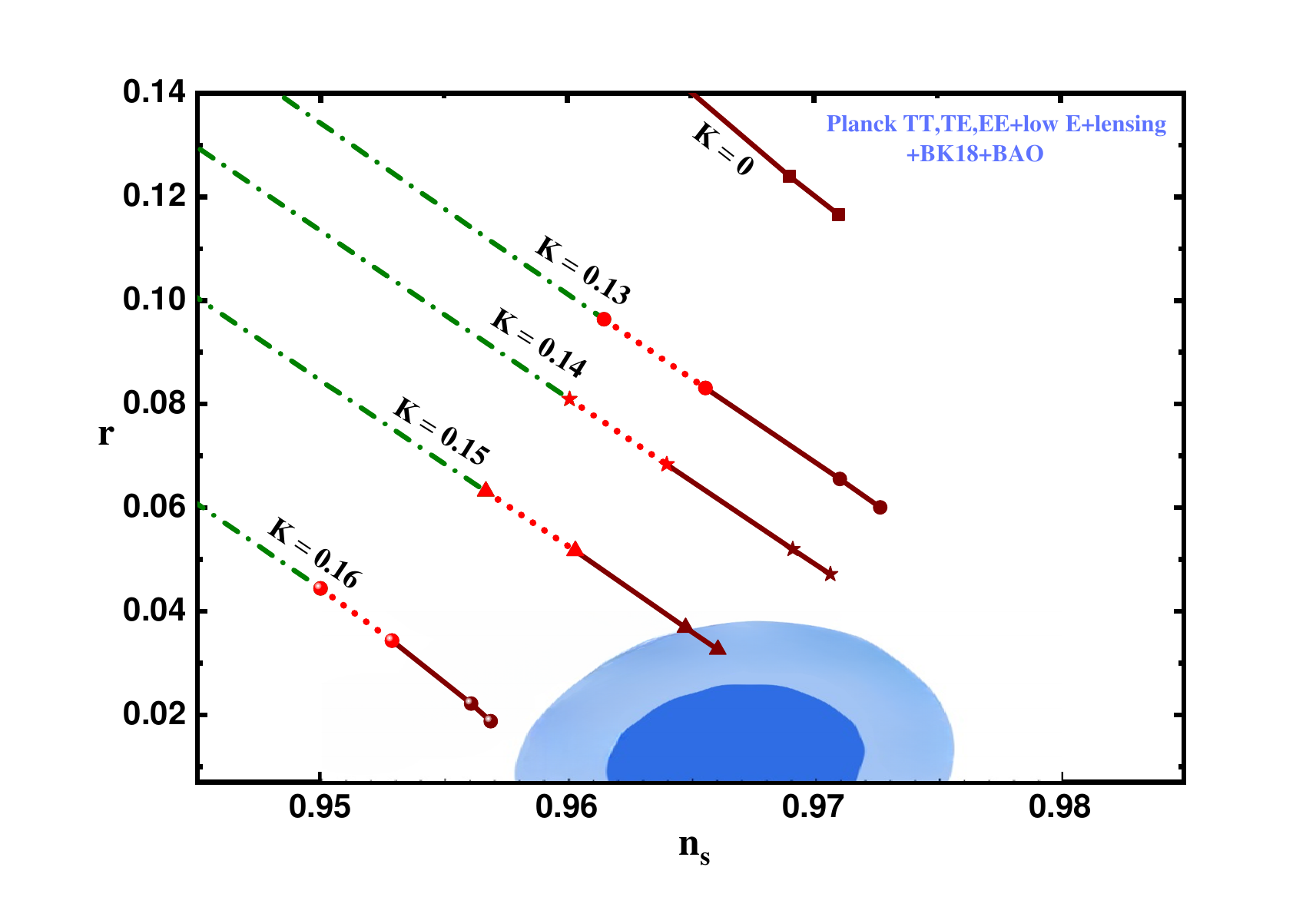}
    \caption{The {r} versus \(n_s\) plots for different values of {K} over a range of \(\overline{\omega }_{\text{re} }\) taken: $\overline{\omega }_{re} < 0$(green), $0 \le \overline{\omega }_{re} \le \frac{1}{6}$(red), $\frac{1}{6} < \overline{\omega }_{re} \le 1$(brown). The dark and light blue shadings corresponds to $1\sigma$ and $2\sigma$ bounds from Planck 2018(TT,TE,EE+lowE+lensing)+BK18+BAO\cite{ade2021improved}.}
\label{F5}
\end{figure}

Defining \(\frac{\phi _{\text{end}}}{M_P}=y\). The relation of field \(\phi\) and \(H\), and the condition for termination of inflation, along with eq. (\ref{36}) gives expression for \(V_{\text{end}}\) in terms of $x$ and $y$ as

\begin{equation} \label{37}
V_{\text{end}}(\phi )=\frac{1}{2}m^2\left(1-\text{K} \ln \left(\frac{\phi _{\text{end}}^2}{M_P^2}\right)\right)\phi _{\text{end}}^2\\
\\
\text{                      }=\frac{ 3H_*^2M_P^2\left(1-\text{K} \ln y^2\right)y^2}{x^2 \left(1-\text{K} \ln x^2\right)}\text{
  }.
\end{equation}
Now, the value of y obtained using the condition for termination of inflation (\(\epsilon =1\)) along with the expressions for $\Delta $\(N_*\), \(H_*\) and \(V_{\text{end} }\) from eqs. (\ref{33}), (\ref{36}) and (\ref{37}) can be inserted in eqs. (\ref{18}) and (\ref{19}) and then these two equations along with eq. (\ref{34}) gives number of reheating e-folds \(N_{\text{re}}\) and reheating temperature \(T_{\text{re}}\). The \(N_{\text{re}}\) and \(T_{\text{re}}\) versus \(n_s\) plots, along with Planck 2018+BK18+BAO bounds, for 4 different K values for this model are presented graphically in figure \ref{F4}.\\
By demanding \(T_{\text{re}}> 100\) GeV for production of weak-scale dark matter and solving eqs. (\ref{18}) and (\ref{34}), the bounds on \(n_s\) are obtained and are reflected on eq. (\ref{33}) and eq. (\ref{35}) to obtain bounds on  $\Delta $\(N_*\) and r. All the obtained bounds for various choices of {K} are shown in table (\ref{T2}). The r versus \(n_s\) plots, along with Planck 2018 (TT,TE,EE+lowE+lensing) and added BK18+BAO constraints, for a range of K values are presented graphically in figure \ref{F5}. The figure \ref{F5} shows that the tensor- to scalar ratio is inside the viable range (\(r<0.036)\) for {K} values closer to  ${\text{K}} = 0.15$. The value $\text{K} = 0$ gives us the normal quadratic chaotic potential.
The range of \(\overline{\omega }_{\text{re}}\) for which our obtained data for ${\text{K}} = 0.15$ is compatible with Planck 2018+BK18+BAO \(2\sigma\) bound on \(n_s\) and  {r} give ($0.7 < \overline{\omega }_{\text{re}} \le 1$).\\
We have also found the viable range of the reheat temperature and number of e-foldings for each case which shows compatibility with Planck 2018+BK18+BAO \(\text{1$\sigma $}\) bound on $n_{s}$ using figure \ref{F4} and the findings have been clearly presented in a tabular format in table \ref{T3}. The table \ref{T3} shows that the curve corresponding to $\overline{\omega }_{\text{re}}=\frac{1}{6}$ for K=0.13, ($\frac{1}{6} \le \overline{\omega }_{\text{re}} \le \frac{2}{3}$) for 0.14 and ($\frac{2}{3} \le \overline{\omega }_{\text{re}} \le 1$) for K=0.15, give every possible value of reheating temperature ($10^{-2}$ GeV to $10^{16}$ GeV) while K=0.16 shows incompatibility with data for all $\overline{\omega }_{\text{re}}$ taken. The \(n_s\) values for these $\overline{\omega }_{\text{re}}$ ranges are 0.966 for K=0.13 and (0.964 $<$ \(n_s\) $\leq $ 0.969) for 0.14 while it is (0.965 $<$ \(n_s\) $\leq $ 0.966) for K=0.15 which sets limit on tensor to scalar ratio(r) and the obtained values of r are 0.083 for K=0.13 and (0.068 $\ge $ r $\ge $ 0.052) for 0.14 while it is (0.037 $\ge $ r $\ge $ 0.033) for K=0.15 and only the r values for {K}=0.15 are satisfying the condition (\(r<0.036)\).
\begin{table}[!t]
\caption{The permissible range for $n_s$, $\Delta N_*$ and $r$ for different K values for modified Quadratic Chaotic inflation by demanding $T_{re} \geq 100GeV$}\label{T2}%
    \centering
    \begin{tabular}{|c|c|c|c|c|c|}
        \hline
         \multirow{5}{*}{K = $0.13$} & Average Equation of state & $n_s$ & $\Delta N_*$ & $r$ \\
        \cline{2-5}
        & $-\frac{1}{3} \le \overline{\omega }_{re} \le 0$ & $0.932 \le n_s \le 0.962$ & $26.07 \le \Delta  N_* \le 47.03$ & $0.195  \ge r \ge 0.096$ \\
         \cline{2-5}
        & $0 \le \overline{\omega }_{re} \le \frac{1}{6}$ & $0.962 \le n_s \le 0.966$ & $47.03 \le \Delta  N_* \le 52.95$ & $0.096  \ge r \ge 0.083$ \\
         \cline{2-5}
        & $\frac{1}{6} \le \overline{\omega }_{re} \le \frac{2}{3}$ & $0.966 \le n_s \le 0.971$ & $52.95 \le \Delta  N_* \le 63.56$ & $0.083  \ge r \ge 0.066$ \\
        \cline{2-5}
        & $\frac{2}{3} \le \overline{\omega }_{re} \le 1$ & $0.971 \le n_s \le 0.973$ & $63.56 \le \Delta  N_* \le 67.68$ & $0.066  \ge r \ge 0.060$ \\
        \hlineB{4}
        \multirow{4}{*}{K = $0.14$}
        & $-\frac{1}{3} \le \overline{\omega }_{re} \le 0$ & $0.931 \le n_s \le 0.960$ & $26.02 \le \Delta  N_* \le 46.94$ & $0.175  \ge r \ge 0.080$ \\
         \cline{2-5}
        & $0 \le \overline{\omega }_{re} \le \frac{1}{6}$ & $0.960 \le n_s \le 0.964$ & $46.94 \le \Delta  N_* \le 52.86$ & $0.080  \ge r \ge 0.068$ \\
         \cline{2-5}
        & $\frac{1}{6} \le \overline{\omega }_{re} \le \frac{2}{3}$ & $0.964 \le n_s \le 0.969$ & $52.86 \le \Delta  N_* \le 63.46$ & $0.068  \ge r \ge 0.052$ \\
        \cline{2-5}
        & $\frac{2}{3} \le \overline{\omega }_{re} \le 1$ & $0.969 \le n_s \le 0.971$ & $63.46 \le \Delta  N_* \le 67.56$ & $0.052  \ge r \ge 0.047$ \\
        \hlineB{4}
         \multirow{4}{*}{K = $0.15$} 
        & $-\frac{1}{3} \le \overline{\omega }_{re} \le 0$ & $0.929 \le n_s \le 0.957$ & $25.95 \le \Delta  N_* \le 46.84$ & $0.152  \ge r \ge 0.063$ \\
         \cline{2-5}
        & $0 \le \overline{\omega }_{re} \le \frac{1}{6}$ & $0.957 \le n_s \le 0.960$ & $46.84 \le \Delta  N_* \le 52.74$ & $0.063  \ge r \ge 0.052$ \\
         \cline{2-5}
        & $\frac{1}{6} \le \overline{\omega }_{re} \le \frac{2}{3}$ & $0.960 \le n_s \le 0.965$ & $52.74 \le \Delta  N_* \le 63.31$ & $0.052  \ge r \ge 0.037$ \\
        \cline{2-5}
        & $\frac{2}{3} \le \overline{\omega }_{re} \le 1$ & $0.965 \le n_s \le 0.966$ & $63.31 \le \Delta  N_* \le 67.40$ & $0.037  \ge r \ge 0.033$ \\
         \hlineB{4}
         \multirow{4}{*}{K = $0.16$} 
        & $-\frac{1}{3} \le \overline{\omega }_{re} \le 0$ & $0.925 \le n_s \le 0.950$ & $25.88 \le \Delta  N_* \le 46.71$ & $0.129  \ge r \ge 0.044$ \\
         \cline{2-5}
        & $0 \le \overline{\omega }_{re} \le \frac{1}{6}$ & $0.950 \le n_s \le 0.953$ & $46.71 \le \Delta  N_* \le 52.60$ & $0.044  \ge r \ge 0.034$ \\
         \cline{2-5}
        & $\frac{1}{6} \le \overline{\omega }_{re} \le \frac{2}{3}$ & $0.953 \le n_s \le 0.956$ & $52.60 \le \Delta  N_* \le 63.11$ & $0.034  \ge r \ge 0.022$ \\
        \cline{2-5}
        & $\frac{2}{3} \le \overline{\omega }_{re} \le 1$ & $0.956 \le n_s \le 0.957$ & $63.11 \le \Delta  N_* \le 67.18$ & $0.022  \ge r \ge 0.019$ \\
        \hline
    \end{tabular}
\end{table}

\begin{table}[!h]
 \caption{The range of $T_{re}$ and $N_{re}$ for potential \(V(\phi )=\frac{1}{2}m^2\left(1-\text{K} \ln \frac{\phi ^2}{M_P^2}\right)\phi^2\) which shows compatibility  with Planck 2018+BK18+BAO} \(\text{1$\sigma $}\) bound on $n_{s}$.\label{T3}
\centering
\scalebox{0.9}{
    \begin{tabular}{|c|c|c|c|c|c|c|}
    \hline
     & \multicolumn{2}{|c|}{K $=0.13$}   &  \multicolumn{2}{|c|}{K $=0.14$} &  \multicolumn{2}{|c|}{K $=0.15$}  \\
    \cline{2-7}
         & $T_{re}${(GeV)} & $N_{re}$ & $T_{re}${(GeV)} & $N_{re}$ & $T_{re}${(GeV)} & $N_{re}$   \\
         \hline
         $\overline{\omega}_{re}=-\frac{1}{3}$ & $T_{re} \ge 1.95\times10^{11}$ & $N_{re} \le 19.53$ & $T_{re} \ge 1.50\times10^{12}$ & $N_{re} \le 15.40$ & $T_{re} \ge 1.12\times10^{15}$ & $N_{re} \le 1.62$ \\
          \hline
           $\overline{\omega}_{re}=0$ & $T_{re} \ge 642.01$ & $N_{re} \le 39.07$ & $T_{re} \ge 3.07 \times10^{5}$ & $N_{re} \le 30.81$ & $T_{re} \ge 2.22\times10^{14}$ & $N_{re} \le 3.24$ \\
          \hline
        $\overline{\omega}_{re}=\frac{1}{6}$ & $10^{16} \ge T_{re} \ge 10^{-2}$ & $N_{re} \le 78.13$ & $10^{16} \ge T_{re} \ge 10^{-2}$ & $N_{re} \le 61.61$ & $T_{re} \ge 8.7\times10^{12}$ & $N_{re} \le 6.48$ \\
          \hline
         $\overline{\omega}_{re}=\frac{2}{3}$ & $T_{re} \ge 211.89$ & $N_{re} \le 24.22$ & $10^{16} \ge T_{re} \ge 10^{-2}$ & $N_{re} \le 45.13$ & $10^{16} \ge T_{re} \ge 10^{-2}$ & $N_{re} \ge 0$ \\
          \hline
          $\overline{\omega}_{re}=1$ & $T_{re} \ge 3.85 \times10^{7}$ & $N_{re} \le 12.11$ & $T_{re} \ge 5.70$ & $N_{re} \le 22.56$ & $10^{16} \ge T_{re} \ge 10^{-2}$ & $N_{re} \le 76.75$ \\
          \hline
    \end{tabular}}
    \label{T3}
\end{table}
\section{Discussion and conclusion}\label{S4}
In this work, we have considered a modified form of quadratic chaotic inflation. Our primary goal is to study the reheating phase in light of Planck 2018+BK18+BAO observations. For that, we have considered two parameters, namely duration of reheating \(N_{\text{re}}\) and reheating temperature \(T_{\text{re}}\) and obtained their variation as function of scalar spectral index \(n_{\text{s}}\) by considering a suitable range of effective equation of state \(\overline{\omega }_{\text{re}}\). By demanding \(T_{\text{re}}> 100\) GeV for production of weak-scale dark matter and allowing \(\overline{\omega }_{\text{re}}\) to vary in the range ($-\frac{1}{3} \le \overline{\omega }_{\text{re}} \le 1$), we tried to find the permissible ranges for \(n_{\text{s}}\), $\Delta  N_*$ and tensor-to-scalar ratio(r) for our models.\\
We first restudied the simple quadratic chaotic inflation using the most recent Planck 2018+BK18+BAO data and found that the condition \(T_{\text{re}}\) $>$ 100 GeV gives ($0 \le \overline{\omega }_{\text{re}} \le 1$) for \(n_s\) to lie inside \(2\sigma\) bounds while if we demand \(n_{s }\) to lie within \(1\sigma\) bounds than the allowed range of \(\overline{\omega }_{\text{re}}\) is ($0.127 \le \overline{\omega }_{\text{re}} \le 1$). Within these ranges of \(\overline{\omega }_{\text{re}}\), r is greater than the observational bound on {r}, i.e. (\(r<0.036)\).\\
Since the normal quadratic chaotic potential is not favoring the observational data. We have considered a modified form of quadratic chaotic potential where a logarithmic correction containing a model parameter K is added to the mass term. We have found that for each value of model parameter K of the modified model, there is only a specific range of inflaton field $(\phi)$  within which the model is defined and the correction part is not dominant over the actual quadratic term of potential. We have constrained ourself to only those regions for our analysis. By imposing the reheating conditions on this model, we found that the constraints on \(n_{\text{s}}\) and r are consistent with Planck's 2018 and BK18 data for {K} values closer to ($ {\text{K}} = 0.15$). The range of \(\overline{\omega }_{\text{re}}\) for which our obtained data is compatible with Planck 2018+BK18+BAO \(2\sigma\) bound on \(n_s\) and  {r} give ($0.7 < \overline{\omega }_{\text{re}} \le 1$) for K=0.15.\\
Also, from the plots showing the variation of \(T_{\text{re}}\) with \(n_s\), we have found that different values of  K and \(\overline{\omega }_{\text{re}}\) give different ranges of reheating temperature as compatible with Planck's \(\text{1$\sigma $}\) bounds on \(n_{\text{s}}\), but if we allow \(T_{\text{re}}\) to vary over the whole range ( $10^{-2}$ GeV to $10^{16}$ GeV) , then \(\overline{\omega }_{\text{re}}\) is restricted to 
 ($0.074 \le \overline{\omega }_{\text{re}} \le 0.570$) for K=0.13, ($0.120 \le \overline{\omega }_{\text{re}} \le 0.854$) for K=0.14 and ($0.307 \le \overline{\omega }_{\text{re}} \le 1$) for K=0.15 while K=0.16 shows incompatibility with \(\text{1$\sigma $}\) bounds on \(n_{\text{s}}\)  for all $\overline{\omega }_{\text{re}}$ taken.\\
To conclude, the reheating study shows that the values of K close to 0.15 are the favorable ones and the \(\overline{\omega }_{\text{re}}\) range satisfying the observational data for K=0.15 suggests the possible production of Feebly Interacting Massive Particle(FIMP) and Weakly Interacting Massive Particle(WIMP)-like dark matter particles{\cite{Haque,2022PhRvD.106b3506H,chakraborty2023inflaton} and primordial black holes \cite{Harada:2013epa}. Elaborated study of possible particle production  will be done in our future publications. The findings of the reheating study prove that even a small correction in mass term can help quadratic chaotic potential to favour the latest cosmological observations. Also, we have found that considering the reheating constraints, the average equation of state parameter \(\overline{\omega }_{\text{re}}\) plays a vital role in defining the compatible range of reheating parameters, which effectively narrows the model's viable parameter space and significantly increases the model's accuracy\\
\section*{Acknowledgments}
 SY would like to acknowledge the Ministry of Education, Government of India, for providing fellowship. UAY acknowledges support from an Institute Chair Professorship of IIT Bombay.
\printbibliography

@article{adshead_inflation_2011,
	title = {Inflation and the scale dependent spectral index: prospects and strategies},
	volume = {2011},
	issn = {1475-7516},
	shorttitle = {Inflation and the scale dependent spectral index},
	doi = {10.1088/1475-7516/2011/02/021},
	number = {02},
	journal = {Journal of Cosmology and Astroparticle Physics},
	author = {Adshead, Peter and Easther, Richard and Pritchard, Jonathan and Loeb, Abraham},
	month = feb,
	year = {2011},
	pages = {021--021},
}

@article{albrecht_reheating_1982,
	title = {Reheating an {Inflationary} {Universe}},
	volume = {48},
	issn = {0031-9007},
	doi = {10.1103/PhysRevLett.48.1437},
	language = {en},
	number = {20},
	journal = {Physical Review Letters},
	author = {Albrecht, Andreas and Steinhardt, Paul J. and Turner, Michael S. and Wilczek, Frank},
	month = may,
	year = {1982},
	pages = {1437--1440},
}

@article{allahverdi_reheating_2010,
	title = {Reheating in {Inflationary} {Cosmology}: {Theory} and {Applications}},
	volume = {60},
	issn = {0163-8998, 1545-4134},
	shorttitle = {Reheating in {Inflationary} {Cosmology}},
	doi = {10.1146/annurev.nucl.012809.104511},
	language = {en},
	number = {1},
	journal = {Annual Review of Nuclear and Particle Science},
	author = {Allahverdi, Rouzbeh and Brandenberger, Robert and Cyr-Racine, Francis-Yan and Mazumdar, Anupam},
	month = nov,
	year = {2010},
	pages = {27--51},
}

@article{ballesteros_radiative_2016,
	title = {Radiative plateau inflation},
	volume = {2016},
	issn = {1029-8479},
	doi = {10.1007/JHEP02(2016)153},
	language = {en},
	number = {2},
	journal = {Journal of High Energy Physics},
	author = {Ballesteros, Guillermo and Tamarit, Carlos},
	month = feb,
	year = {2016},
	pages = {153},
}

@article{boubekeur_current_2015,
	title = {Do current data prefer a nonminimally coupled inflaton?},
	volume = {91},
	issn = {1550-7998, 1550-2368},
	doi = {10.1103/PhysRevD.91.103004},
	language = {en},
	number = {10},
	journal = {Physical Review D},
	author = {Boubekeur, Lotfi and Giusarma, Elena and Mena, Olga and Ramírez, Héctor},
	month = may,
	year = {2015},
	pages = {103004},
}

@article{boyanovsky_preheating_1996,
	title = {Preheating and {Reheating} in {Inflationary} {Cosmology}: a pedagogical survey},
	copyright = {Assumed arXiv.org perpetual, non-exclusive license to distribute this article for submissions made before January 2004},
	shorttitle = {Preheating and {Reheating} in {Inflationary} {Cosmology}},
	doi = {10.48550/ARXIV.ASTRO-PH/9609007},
	author = {Boyanovsky, D. and de Vega, H. J. and Holman, R. and Salgado, J. F. J.},
	year = {1996},
}

@article{cook_reheating_2015,
	title = {Reheating predictions in single field inflation},
	volume = {2015},
	issn = {1475-7516},
	doi = {10.1088/1475-7516/2015/04/047},
	number = {04},
	journal = {Journal of Cosmology and Astroparticle Physics},
	author = {Cook, Jessica L. and Dimastrogiovanni, Emanuela and Easson, Damien A. and Krauss, Lawrence M.},
	month = apr,
	year = {2015},
	pages = {047--047},
}

@article{dai_reheating_2014,
	title = {Reheating {Constraints} to {Inflationary} {Models}},
	volume = {113},
	issn = {0031-9007, 1079-7114},

	doi = {10.1103/PhysRevLett.113.041302},
	language = {en},
	number = {4},

	journal = {Physical Review Letters},
	author = {Dai, Liang and Kamionkowski, Marc and Wang, Junpu},
	month = jul,
	year = {2014},
	pages = {041302},
}

@article{desroche_preheating_2005,
	title = {Preheating in new inflation},
	volume = {71},
	issn = {1550-7998, 1550-2368},

	doi = {10.1103/PhysRevD.71.103516},
	language = {en},
	number = {10},

	journal = {Physical Review D},
	author = {Desroche, Mariel and Felder, Gary N. and Kratochvil, Jan M. and Linde, Andrei},
	month = may,
	year = {2005},
	pages = {103516},
}

@article{drewes_kinematics_2013,
	title = {The kinematics of cosmic reheating},
	volume = {875},
	issn = {05503213},

	doi = {10.1016/j.nuclphysb.2013.07.009},
	language = {en},
	number = {2},

	journal = {Nuclear Physics B},
	author = {Drewes, Marco and Kang, Jin U},
	month = oct,
	year = {2013},
	pages = {315--350},
}

@article{dunkley_five-year_2009,
	title = {{FIVE}-{YEAR} \textit{{WILKINSON} {MICROWAVE} {ANISOTROPY} {PROBE}} {OBSERVATIONS}: {LIKELIHOODS} {AND} {PARAMETERS} {FROM} {THE} \textit{{WMAP}} {DATA}},
	volume = {180},
	issn = {0067-0049, 1538-4365},
	shorttitle = {{FIVE}-{YEAR} \textit{{WILKINSON} {MICROWAVE} {ANISOTROPY} {PROBE}} {OBSERVATIONS}},

	doi = {10.1088/0067-0049/180/2/306},
	number = {2},

	journal = {The Astrophysical Journal Supplement Series},
	author = {Dunkley, J. and Komatsu, E. and Nolta, M. R. and Spergel, D. N. and Larson, D. and Hinshaw, G. and Page, L. and Bennett, C. L. and Gold, B. and Jarosik, N. and Weiland, J. L. and Halpern, M. and Hill, R. S. and Kogut, A. and Limon, M. and Meyer, S. S. and Tucker, G. S. and Wollack, E. and Wright, E. L.},
	month = feb,
	year = {2009},
	pages = {306--329},
}

@article{enqvist_does_2014,
	title = {Does {Planck} really rule out monomial inflation?},
	volume = {2014},
	issn = {1475-7516},

	doi = {10.1088/1475-7516/2014/02/034},
	number = {02},

	journal = {Journal of Cosmology and Astroparticle Physics},
	author = {Enqvist, Kari and Karčiauskas, Mindaugas},
	month = feb,
	year = {2014},
	pages = {034--034},
}

@article{felder_instant_1998,
	title = {Instant {Preheating}},
	copyright = {Assumed arXiv.org perpetual, non-exclusive license to distribute this article for submissions made before January 2004},

	doi = {10.48550/ARXIV.HEP-PH/9812289},

	author = {Felder, Gary and Kofman, Lev and Linde, Andrei},
	year = {1998},
}

@article{giudice_cosmological_2001,
	title = {The cosmological moduli problem and preheating},
	volume = {2001},
	issn = {1029-8479},

	doi = {10.1088/1126-6708/2001/06/020},
	number = {06},

	journal = {Journal of High Energy Physics},
	author = {Giudice, Gian Francesco and Tkachev, Igor I and Riotto, Antonio},
	month = jun,
	year = {2001},
	pages = {020--020},
}

@article{goswami_reconciling_2018,
	title = {Reconciling low multipole anomalies and reheating in single field inflationary models},
	volume = {2018},
	issn = {1475-7516},

	doi = {10.1088/1475-7516/2018/10/018},
	number = {10},

	journal = {Journal of Cosmology and Astroparticle Physics},
	author = {Goswami, Rajesh and Yajnik, Urjit A.},
	month = oct,
	year = {2018},
	pages = {018--018},
}

@article{goswami_reheating_2020,
	title = {Reheating constraints to modulus mass for single field inflationary models},
	volume = {960},
	issn = {05503213},

	doi = {10.1016/j.nuclphysb.2020.115211},
	language = {en},

	journal = {Nuclear Physics B},
	author = {Goswami, Rajesh and Yajnik, Urjit A.},
	month = nov,
	year = {2020},
	pages = {115211},
}

@article{guth_inflationary_1981,
	title = {Inflationary universe: {A} possible solution to the horizon and flatness problems},
	volume = {23},
	shorttitle = {Inflationary universe},

	doi = {10.1103/PhysRevD.23.347},
	number = {2},

	journal = {Physical Review D},
	author = {Guth, Alan H.},
	month = jan,
	year = {1981},
	pages = {347--356},
}

@article{guth_quantum_1985,
	title = {Quantum mechanics of the scalar field in the new inflationary universe},
	volume = {32},
	issn = {0556-2821},

	doi = {10.1103/PhysRevD.32.1899},
	language = {en},
	number = {8},

	journal = {Physical Review D},
	author = {Guth, Alan H. and Pi, So-Young},
	month = oct,
	year = {1985},
	pages = {1899--1920},
}

@article{kannike_dynamically_2015,
	title = {Dynamically induced {Planck} scale and inflation},
	volume = {2015},
	issn = {1029-8479},

	doi = {10.1007/JHEP05(2015)065},
	language = {en},
	number = {5},

	journal = {Journal of High Energy Physics},
	author = {Kannike, Kristjan and Hütsi, Gert and Pizza, Liberato and Racioppi, Antonio and Raidal, Martti and Salvio, Alberto and Strumia, Alessandro},
	month = may,
	year = {2015},
	pages = {65},
}

@article{kasuya_quadratic_2018,
	title = {Quadratic chaotic inflation with a logarithmic-corrected mass},
	volume = {98},
	issn = {2470-0010, 2470-0029},

	doi = {10.1103/PhysRevD.98.123515},
	language = {en},
	number = {12},

	journal = {Physical Review D},
	author = {Kasuya, Shinta and Taira, Mayuko},
	month = dec,
	year = {2018},
	pages = {123515},
}

@article{kofman_reheating_1994,
	title = {Reheating after {Inflation}},
	volume = {73},
	issn = {0031-9007},

	doi = {10.1103/PhysRevLett.73.3195},
	language = {en},
	number = {24},

	journal = {Physical Review Letters},
	author = {Kofman, Lev and Linde, Andrei and Starobinsky, Alexei A.},
	month = dec,
	year = {1994},
	pages = {3195--3198},
}

@article{kofman_reheating_1998,
	title = {Reheating and {Preheating} after {Inflation}},
	copyright = {Assumed arXiv.org perpetual, non-exclusive license to distribute this article for submissions made before January 2004},

	doi = {10.48550/ARXIV.HEP-PH/9802285},

	author = {Kofman, Lev},
	year = {1998},
}

@article{kofman_towards_1997,
	title = {Towards the theory of reheating after inflation},
	volume = {56},
	issn = {0556-2821, 1089-4918},

	doi = {10.1103/PhysRevD.56.3258},
	language = {en},
	number = {6},

	journal = {Physical Review D},
	author = {Kofman, Lev and Linde, Andrei and Starobinsky, Alexei A.},
	month = sep,
	year = {1997},
	pages = {3258--3295},
}

@article{komatsu_seven-year_2011,
	title = {{SEVEN}-{YEAR} \textit{{WILKINSON} {MICROWAVE} {ANISOTROPY} {PROBE}} ( \textit{{WMAP}} ) {OBSERVATIONS}: {COSMOLOGICAL} {INTERPRETATION}},
	volume = {192},
	issn = {0067-0049, 1538-4365},
	shorttitle = {{SEVEN}-{YEAR} \textit{{WILKINSON} {MICROWAVE} {ANISOTROPY} {PROBE}} ( \textit{{WMAP}} ) {OBSERVATIONS}},

	doi = {10.1088/0067-0049/192/2/18},
	number = {2},

	journal = {The Astrophysical Journal Supplement Series},
	author = {Komatsu, E. and Smith, K. M. and Dunkley, J. and Bennett, C. L. and Gold, B. and Hinshaw, G. and Jarosik, N. and Larson, D. and Nolta, M. R. and Page, L. and Spergel, D. N. and Halpern, M. and Hill, R. S. and Kogut, A. and Limon, M. and Meyer, S. S. and Odegard, N. and Tucker, G. S. and Weiland, J. L. and Wollack, E. and Wright, E. L.},
	month = feb,
	year = {2011},
	pages = {18},
}

@article{linde_new_1982,
	title = {A new inflationary universe scenario: {A} possible solution of the horizon, flatness, homogeneity, isotropy and primordial monopole problems},
	volume = {108},
	issn = {03702693},
	shorttitle = {A new inflationary universe scenario},

	doi = {10.1016/0370-2693(82)91219-9},
	language = {en},
	number = {6},

	journal = {Physics Letters B},
	author = {Linde, A.D.},
	month = feb,
	year = {1982},
	pages = {389--393},
}

@article{linde_chaotic_1983,
	title = {Chaotic inflation},
	volume = {129},
	issn = {03702693},

	doi = {10.1016/0370-2693(83)90837-7},
	language = {en},
	number = {3-4},

	journal = {Physics Letters B},
	author = {Linde, A.D.},
	month = sep,
	year = {1983},
	pages = {177--181},
}

@misc{martin_encyclopaedia_2013,
	title = {Encyclopaedia {Inflationaris}},

	doi = {10.48550/arXiv.1303.3787},

	author = {Martin, Jerome and Ringeval, Christophe and Vennin, Vincent},
	month = sep,
	year = {2013},
	note = {arXiv:1303.3787 [astro-ph, physics:gr-qc, physics:hep-ph, physics:hep-th]},
}

@article{martin_inflation_2003,
	title = {Inflation and {Precision} {Cosmology}},
	copyright = {Assumed arXiv.org perpetual, non-exclusive license to distribute this article for submissions made before January 2004},

	doi = {10.48550/ARXIV.ASTRO-PH/0312492},

	author = {Martin, Jerome},
	year = {2003},
}

@article{martin_observing_2015,
	title = {Observing {Inflationary} {Reheating}},
	volume = {114},
	issn = {0031-9007, 1079-7114},

	doi = {10.1103/PhysRevLett.114.081303},
	language = {en},
	number = {8},

	journal = {Physical Review Letters},
	author = {Martin, Jérôme and Ringeval, Christophe and Vennin, Vincent},
	month = feb,
	year = {2015},
	pages = {081303},
}

@article{martin_first_2010,
	title = {First {CMB} constraints on the inflationary reheating temperature},
	volume = {82},
	issn = {1550-7998, 1550-2368},

	doi = {10.1103/PhysRevD.82.023511},
	language = {en},
	number = {2},

	journal = {Physical Review D},
	author = {Martin, Jérôme and Ringeval, Christophe},
	month = jul,
	year = {2010},
	pages = {023511},
}

@article{martin_inflation_2006,
	title = {Inflation after {WMAP3}: confronting the slow-roll and exact power spectra with {CMB} data},
	volume = {2006},
	issn = {1475-7516},
	shorttitle = {Inflation after {WMAP3}},

	doi = {10.1088/1475-7516/2006/08/009},
	number = {08},

	journal = {Journal of Cosmology and Astroparticle Physics},
	author = {Martin, Jérôme and Ringeval, Christophe},
	month = aug,
	year = {2006},
	pages = {009--009},
}

@article{marzola_minimal_2016,
	title = {Minimal but non-minimal inflation and electroweak symmetry breaking},
	volume = {2016},
	issn = {1475-7516},

	doi = {10.1088/1475-7516/2016/10/010},
	number = {10},

	journal = {Journal of Cosmology and Astroparticle Physics},
	author = {Marzola, Luca and Racioppi, Antonio},
	month = oct,
	year = {2016},
	pages = {010--010},
}

@article{mielczarek_reheating_2011,
	title = {Reheating temperature from the {CMB}},
	volume = {83},
	issn = {1550-7998, 1550-2368},

	doi = {10.1103/PhysRevD.83.023502},
	language = {en},
	number = {2},

	journal = {Physical Review D},
	author = {Mielczarek, Jakub},
	month = jan,
	year = {2011},
	pages = {023502},
}

@article{mukhanov_quantum_1981,
	title = {Quantum fluctuations and a nonsingular universe},
	volume = {33},

	journal = {ZhETF Pisma Redaktsiiu},
	author = {Mukhanov, V. F. and Chibisov, G. V.},
	month = may,
	year = {1981},
	note = {ADS Bibcode: 1981ZhPmR..33..549M},
	pages = {549--553},
}

@article{nakayama_polynomial_2013,
	title = {Polynomial chaotic inflation in the {Planck} era},
	volume = {725},
	issn = {03702693},

	doi = {10.1016/j.physletb.2013.06.050},
	language = {en},
	number = {1-3},

	journal = {Physics Letters B},
	author = {Nakayama, Kazunori and Takahashi, Fuminobu and Yanagida, Tsutomu T.},
	month = aug,
	year = {2013},
	pages = {111--114},
}

@article{nakayama_running_2010,
	title = {Running kinetic inflation},
	volume = {2010},
	issn = {1475-7516},

	doi = {10.1088/1475-7516/2010/11/009},
	number = {11},

	journal = {Journal of Cosmology and Astroparticle Physics},
	author = {Nakayama, Kazunori and Takahashi, Fuminobu},
	month = nov,
	year = {2010},
	pages = {009--009},
}

@article{pallis_kinetically_2015,
	title = {Kinetically modified nonminimal chaotic inflation},
	volume = {91},
	issn = {1550-7998, 1550-2368},

	doi = {10.1103/PhysRevD.91.123508},
	language = {en},
	number = {12},

	journal = {Physical Review D},
	author = {Pallis, Constantinos},
	month = jun,
	year = {2015},
	pages = {123508},
}

@article{ade_planck_2014,
	title = {\textit{{Planck}} 2013 results. {XVI}. {Cosmological} parameters},
	volume = {571},
	issn = {0004-6361, 1432-0746},

	doi = {10.1051/0004-6361/201321591},

	journal = {Astronomy \& Astrophysics},
	author = {Ade, P. A. R. and Aghanim, N. and Armitage-Caplan, C. and Arnaud, M. and Ashdown, M. and Atrio-Barandela, F. and Aumont, J. and Baccigalupi, C. and Banday, A. J. and Barreiro, R. B. and Bartlett, J. G. and Battaner, E. and Benabed, K. and Benoît, A. and Benoit-Lévy, A. and Bernard, J.-P. and Bersanelli, M. and Bielewicz, P. and Bobin, J. and Bock, J. J. and Bonaldi, A. and Bond, J. R. and Borrill, J. and Bouchet, F. R. and Bridges, M. and Bucher, M. and Burigana, C. and Butler, R. C. and Calabrese, E. and Cappellini, B. and Cardoso, J.-F. and Catalano, A. and Challinor, A. and Chamballu, A. and Chary, R.-R. and Chen, X. and Chiang, H. C. and Chiang, L.-Y and Christensen, P. R. and Church, S. and Clements, D. L. and Colombi, S. and Colombo, L. P. L. and Couchot, F. and Coulais, A. and Crill, B. P. and Curto, A. and Cuttaia, F. and Danese, L. and Davies, R. D. and Davis, R. J. and De Bernardis, P. and De Rosa, A. and De Zotti, G. and Delabrouille, J. and Delouis, J.-M. and Désert, F.-X. and Dickinson, C. and Diego, J. M. and Dolag, K. and Dole, H. and Donzelli, S. and Doré, O. and Douspis, M. and Dunkley, J. and Dupac, X. and Efstathiou, G. and Elsner, F. and Enßlin, T. A. and Eriksen, H. K. and Finelli, F. and Forni, O. and Frailis, M. and Fraisse, A. A. and Franceschi, E. and Gaier, T. C. and Galeotta, S. and Galli, S. and Ganga, K. and Giard, M. and Giardino, G. and Giraud-Héraud, Y. and Gjerløw, E. and González-Nuevo, J. and Górski, K. M. and Gratton, S. and Gregorio, A. and Gruppuso, A. and Gudmundsson, J. E. and Haissinski, J. and Hamann, J. and Hansen, F. K. and Hanson, D. and Harrison, D. and Henrot-Versillé, S. and Hernández-Monteagudo, C. and Herranz, D. and Hildebrandt, S. R. and Hivon, E. and Hobson, M. and Holmes, W. A. and Hornstrup, A. and Hou, Z. and Hovest, W. and Huffenberger, K. M. and Jaffe, A. H. and Jaffe, T. R. and Jewell, J. and Jones, W. C. and Juvela, M. and Keihänen, E. and Keskitalo, R. and Kisner, T. S. and Kneissl, R. and Knoche, J. and Knox, L. and Kunz, M. and Kurki-Suonio, H. and Lagache, G. and Lähteenmäki, A. and Lamarre, J.-M. and Lasenby, A. and Lattanzi, M. and Laureijs, R. J. and Lawrence, C. R. and Leach, S. and Leahy, J. P. and Leonardi, R. and León-Tavares, J. and Lesgourgues, J. and Lewis, A. and Liguori, M. and Lilje, P. B. and Linden-Vørnle, M. and López-Caniego, M. and Lubin, P. M. and Macías-Pérez, J. F. and Maffei, B. and Maino, D. and Mandolesi, N. and Maris, M. and Marshall, D. J. and Martin, P. G. and Martínez-González, E. and Masi, S. and Massardi, M. and Matarrese, S. and Matthai, F. and Mazzotta, P. and Meinhold, P. R. and Melchiorri, A. and Melin, J.-B. and Mendes, L. and Menegoni, E. and Mennella, A. and Migliaccio, M. and Millea, M. and Mitra, S. and Miville-Deschênes, M.-A. and Moneti, A. and Montier, L. and Morgante, G. and Mortlock, D. and Moss, A. and Munshi, D. and Murphy, J. A. and Naselsky, P. and Nati, F. and Natoli, P. and Netterfield, C. B. and Nørgaard-Nielsen, H. U. and Noviello, F. and Novikov, D. and Novikov, I. and O’Dwyer, I. J. and Osborne, S. and Oxborrow, C. A. and Paci, F. and Pagano, L. and Pajot, F. and Paladini, R. and Paoletti, D. and Partridge, B. and Pasian, F. and Patanchon, G. and Pearson, D. and Pearson, T. J. and Peiris, H. V. and Perdereau, O. and Perotto, L. and Perrotta, F. and Pettorino, V. and Piacentini, F. and Piat, M. and Pierpaoli, E. and Pietrobon, D. and Plaszczynski, S. and Platania, P. and Pointecouteau, E. and Polenta, G. and Ponthieu, N. and Popa, L. and Poutanen, T. and Pratt, G. W. and Prézeau, G. and Prunet, S. and Puget, J.-L. and Rachen, J. P. and Reach, W. T. and Rebolo, R. and Reinecke, M. and Remazeilles, M. and Renault, C. and Ricciardi, S. and Riller, T. and Ristorcelli, I. and Rocha, G. and Rosset, C. and Roudier, G. and Rowan-Robinson, M. and Rubiño-Martín, J. A. and Rusholme, B. and Sandri, M. and Santos, D. and Savelainen, M. and Savini, G. and Scott, D. and Seiffert, M. D. and Shellard, E. P. S. and Spencer, L. D. and Starck, J.-L. and Stolyarov, V. and Stompor, R. and Sudiwala, R. and Sunyaev, R. and Sureau, F. and Sutton, D. and Suur-Uski, A.-S. and Sygnet, J.-F. and Tauber, J. A. and Tavagnacco, D. and Terenzi, L. and Toffolatti, L. and Tomasi, M. and Tristram, M. and Tucci, M. and Tuovinen, J. and Türler, M. and Umana, G. and Valenziano, L. and Valiviita, J. and Van Tent, B. and Vielva, P. and Villa, F. and Vittorio, N. and Wade, L. A. and Wandelt, B. D. and Wehus, I. K. and White, M. and White, S. D. M. and Wilkinson, A. and Yvon, D. and Zacchei, A. and Zonca, A.},
	month = nov,
	year = {2014},
	pages = {A16},
}

@article{ade_planck_2014-1,
	title = {\textit{{Planck}} 2013 results. {XXII}. {Constraints} on inflation},
	volume = {571},
	issn = {0004-6361, 1432-0746},

	doi = {10.1051/0004-6361/201321569},

	journal = {Astronomy \& Astrophysics},
	author = {Ade, P. A. R. and Aghanim, N. and Armitage-Caplan, C. and Arnaud, M. and Ashdown, M. and Atrio-Barandela, F. and Aumont, J. and Baccigalupi, C. and Banday, A. J. and Barreiro, R. B. and Bartlett, J. G. and Bartolo, N. and Battaner, E. and Benabed, K. and Benoît, A. and Benoit-Lévy, A. and Bernard, J.-P. and Bersanelli, M. and Bielewicz, P. and Bobin, J. and Bock, J. J. and Bonaldi, A. and Bond, J. R. and Borrill, J. and Bouchet, F. R. and Bridges, M. and Bucher, M. and Burigana, C. and Butler, R. C. and Calabrese, E. and Cardoso, J.-F. and Catalano, A. and Challinor, A. and Chamballu, A. and Chiang, H. C. and Chiang, L.-Y. and Christensen, P. R. and Church, S. and Clements, D. L. and Colombi, S. and Colombo, L. P. L. and Couchot, F. and Coulais, A. and Crill, B. P. and Curto, A. and Cuttaia, F. and Danese, L. and Davies, R. D. and Davis, R. J. and De Bernardis, P. and De Rosa, A. and De Zotti, G. and Delabrouille, J. and Delouis, J.-M. and Désert, F.-X. and Dickinson, C. and Diego, J. M. and Dole, H. and Donzelli, S. and Doré, O. and Douspis, M. and Dunkley, J. and Dupac, X. and Efstathiou, G. and Enßlin, T. A. and Eriksen, H. K. and Finelli, F. and Forni, O. and Frailis, M. and Franceschi, E. and Galeotta, S. and Ganga, K. and Gauthier, C. and Giard, M. and Giardino, G. and Giraud-Héraud, Y. and González-Nuevo, J. and Górski, K. M. and Gratton, S. and Gregorio, A. and Gruppuso, A. and Hamann, J. and Hansen, F. K. and Hanson, D. and Harrison, D. and Henrot-Versillé, S. and Hernández-Monteagudo, C. and Herranz, D. and Hildebrandt, S. R. and Hivon, E. and Hobson, M. and Holmes, W. A. and Hornstrup, A. and Hovest, W. and Huffenberger, K. M. and Jaffe, A. H. and Jaffe, T. R. and Jones, W. C. and Juvela, M. and Keihänen, E. and Keskitalo, R. and Kisner, T. S. and Kneissl, R. and Knoche, J. and Knox, L. and Kunz, M. and Kurki-Suonio, H. and Lagache, G. and Lähteenmäki, A. and Lamarre, J.-M. and Lasenby, A. and Laureijs, R. J. and Lawrence, C. R. and Leach, S. and Leahy, J. P. and Leonardi, R. and Lesgourgues, J. and Lewis, A. and Liguori, M. and Lilje, P. B. and Linden-Vørnle, M. and López-Caniego, M. and Lubin, P. M. and Macías-Pérez, J. F. and Maffei, B. and Maino, D. and Mandolesi, N. and Maris, M. and Marshall, D. J. and Martin, P. G. and Martínez-González, E. and Masi, S. and Massardi, M. and Matarrese, S. and Matthai, F. and Mazzotta, P. and Meinhold, P. R. and Melchiorri, A. and Mendes, L. and Mennella, A. and Migliaccio, M. and Mitra, S. and Miville-Deschênes, M.-A. and Moneti, A. and Montier, L. and Morgante, G. and Mortlock, D. and Moss, A. and Munshi, D. and Murphy, J. A. and Naselsky, P. and Nati, F. and Natoli, P. and Netterfield, C. B. and Nørgaard-Nielsen, H. U. and Noviello, F. and Novikov, D. and Novikov, I. and O’Dwyer, I. J. and Osborne, S. and Oxborrow, C. A. and Paci, F. and Pagano, L. and Pajot, F. and Paladini, R. and Pandolfi, S. and Paoletti, D. and Partridge, B. and Pasian, F. and Patanchon, G. and Peiris, H. V. and Perdereau, O. and Perotto, L. and Perrotta, F. and Piacentini, F. and Piat, M. and Pierpaoli, E. and Pietrobon, D. and Plaszczynski, S. and Pointecouteau, E. and Polenta, G. and Ponthieu, N. and Popa, L. and Poutanen, T. and Pratt, G. W. and Prézeau, G. and Prunet, S. and Puget, J.-L. and Rachen, J. P. and Rebolo, R. and Reinecke, M. and Remazeilles, M. and Renault, C. and Ricciardi, S. and Riller, T. and Ristorcelli, I. and Rocha, G. and Rosset, C. and Roudier, G. and Rowan-Robinson, M. and Rubiño-Martín, J. A. and Rusholme, B. and Sandri, M. and Santos, D. and Savelainen, M. and Savini, G. and Scott, D. and Seiffert, M. D. and Shellard, E. P. S. and Spencer, L. D. and Starck, J.-L. and Stolyarov, V. and Stompor, R. and Sudiwala, R. and Sunyaev, R. and Sureau, F. and Sutton, D. and Suur-Uski, A.-S. and Sygnet, J.-F. and Tauber, J. A. and Tavagnacco, D. and Terenzi, L. and Toffolatti, L. and Tomasi, M. and Tréguer-Goudineau, J. and Tristram, M. and Tucci, M. and Tuovinen, J. and Valenziano, L. and Valiviita, J. and Van Tent, B. and Varis, J. and Vielva, P. and Villa, F. and Vittorio, N. and Wade, L. A. and Wandelt, B. D. and White, M. and Wilkinson, A. and Yvon, D. and Zacchei, A. and Zibin, J. P. and Zonca, A.},
	month = nov,
	year = {2014},
	pages = {A22},
}

@article{ade_planck_2016,
	title = {\textit{{Planck}} 2015 results: {XIII}. {Cosmological} parameters},
	volume = {594},
	issn = {0004-6361, 1432-0746},
	shorttitle = {\textit{{Planck}} 2015 results},

	doi = {10.1051/0004-6361/201525830},

	journal = {Astronomy \& Astrophysics},
	author = {Ade, P. A. R. and Aghanim, N. and Arnaud, M. and Ashdown, M. and Aumont, J. and Baccigalupi, C. and Banday, A. J. and Barreiro, R. B. and Bartlett, J. G. and Bartolo, N. and Battaner, E. and Battye, R. and Benabed, K. and Benoît, A. and Benoit-Lévy, A. and Bernard, J.-P. and Bersanelli, M. and Bielewicz, P. and Bock, J. J. and Bonaldi, A. and Bonavera, L. and Bond, J. R. and Borrill, J. and Bouchet, F. R. and Boulanger, F. and Bucher, M. and Burigana, C. and Butler, R. C. and Calabrese, E. and Cardoso, J.-F. and Catalano, A. and Challinor, A. and Chamballu, A. and Chary, R.-R. and Chiang, H. C. and Chluba, J. and Christensen, P. R. and Church, S. and Clements, D. L. and Colombi, S. and Colombo, L. P. L. and Combet, C. and Coulais, A. and Crill, B. P. and Curto, A. and Cuttaia, F. and Danese, L. and Davies, R. D. and Davis, R. J. and De Bernardis, P. and De Rosa, A. and De Zotti, G. and Delabrouille, J. and Désert, F.-X. and Di Valentino, E. and Dickinson, C. and Diego, J. M. and Dolag, K. and Dole, H. and Donzelli, S. and Doré, O. and Douspis, M. and Ducout, A. and Dunkley, J. and Dupac, X. and Efstathiou, G. and Elsner, F. and Enßlin, T. A. and Eriksen, H. K. and Farhang, M. and Fergusson, J. and Finelli, F. and Forni, O. and Frailis, M. and Fraisse, A. A. and Franceschi, E. and Frejsel, A. and Galeotta, S. and Galli, S. and Ganga, K. and Gauthier, C. and Gerbino, M. and Ghosh, T. and Giard, M. and Giraud-Héraud, Y. and Giusarma, E. and Gjerløw, E. and González-Nuevo, J. and Górski, K. M. and Gratton, S. and Gregorio, A. and Gruppuso, A. and Gudmundsson, J. E. and Hamann, J. and Hansen, F. K. and Hanson, D. and Harrison, D. L. and Helou, G. and Henrot-Versillé, S. and Hernández-Monteagudo, C. and Herranz, D. and Hildebrandt, S. R. and Hivon, E. and Hobson, M. and Holmes, W. A. and Hornstrup, A. and Hovest, W. and Huang, Z. and Huffenberger, K. M. and Hurier, G. and Jaffe, A. H. and Jaffe, T. R. and Jones, W. C. and Juvela, M. and Keihänen, E. and Keskitalo, R. and Kisner, T. S. and Kneissl, R. and Knoche, J. and Knox, L. and Kunz, M. and Kurki-Suonio, H. and Lagache, G. and Lähteenmäki, A. and Lamarre, J.-M. and Lasenby, A. and Lattanzi, M. and Lawrence, C. R. and Leahy, J. P. and Leonardi, R. and Lesgourgues, J. and Levrier, F. and Lewis, A. and Liguori, M. and Lilje, P. B. and Linden-Vørnle, M. and López-Caniego, M. and Lubin, P. M. and Macías-Pérez, J. F. and Maggio, G. and Maino, D. and Mandolesi, N. and Mangilli, A. and Marchini, A. and Maris, M. and Martin, P. G. and Martinelli, M. and Martínez-González, E. and Masi, S. and Matarrese, S. and McGehee, P. and Meinhold, P. R. and Melchiorri, A. and Melin, J.-B. and Mendes, L. and Mennella, A. and Migliaccio, M. and Millea, M. and Mitra, S. and Miville-Deschênes, M.-A. and Moneti, A. and Montier, L. and Morgante, G. and Mortlock, D. and Moss, A. and Munshi, D. and Murphy, J. A. and Naselsky, P. and Nati, F. and Natoli, P. and Netterfield, C. B. and Nørgaard-Nielsen, H. U. and Noviello, F. and Novikov, D. and Novikov, I. and Oxborrow, C. A. and Paci, F. and Pagano, L. and Pajot, F. and Paladini, R. and Paoletti, D. and Partridge, B. and Pasian, F. and Patanchon, G. and Pearson, T. J. and Perdereau, O. and Perotto, L. and Perrotta, F. and Pettorino, V. and Piacentini, F. and Piat, M. and Pierpaoli, E. and Pietrobon, D. and Plaszczynski, S. and Pointecouteau, E. and Polenta, G. and Popa, L. and Pratt, G. W. and Prézeau, G. and Prunet, S. and Puget, J.-L. and Rachen, J. P. and Reach, W. T. and Rebolo, R. and Reinecke, M. and Remazeilles, M. and Renault, C. and Renzi, A. and Ristorcelli, I. and Rocha, G. and Rosset, C. and Rossetti, M. and Roudier, G. and Rouillé d’Orfeuil, B. and Rowan-Robinson, M. and Rubiño-Martín, J. A. and Rusholme, B. and Said, N. and Salvatelli, V. and Salvati, L. and Sandri, M. and Santos, D. and Savelainen, M. and Savini, G. and Scott, D. and Seiffert, M. D. and Serra, P. and Shellard, E. P. S. and Spencer, L. D. and Spinelli, M. and Stolyarov, V. and Stompor, R. and Sudiwala, R. and Sunyaev, R. and Sutton, D. and Suur-Uski, A.-S. and Sygnet, J.-F. and Tauber, J. A. and Terenzi, L. and Toffolatti, L. and Tomasi, M. and Tristram, M. and Trombetti, T. and Tucci, M. and Tuovinen, J. and Türler, M. and Umana, G. and Valenziano, L. and Valiviita, J. and Van Tent, F. and Vielva, P. and Villa, F. and Wade, L. A. and Wandelt, B. D. and Wehus, I. K. and White, M. and White, S. D. M. and Wilkinson, A. and Yvon, D. and Zacchei, A. and Zonca, A.},
	month = oct,
	year = {2016},
	pages = {A13},
}

@article{ade_planck_2016-1,
	title = {\textit{{Planck}} 2015 results: {XX}. {Constraints} on inflation},
	volume = {594},
	issn = {0004-6361, 1432-0746},
	shorttitle = {\textit{{Planck}} 2015 results},

	doi = {10.1051/0004-6361/201525898},

	journal = {Astronomy \& Astrophysics},
	author = {Ade, P. A. R. and Aghanim, N. and Arnaud, M. and Arroja, F. and Ashdown, M. and Aumont, J. and Baccigalupi, C. and Ballardini, M. and Banday, A. J. and Barreiro, R. B. and Bartolo, N. and Battaner, E. and Benabed, K. and Benoît, A. and Benoit-Lévy, A. and Bernard, J.-P. and Bersanelli, M. and Bielewicz, P. and Bock, J. J. and Bonaldi, A. and Bonavera, L. and Bond, J. R. and Borrill, J. and Bouchet, F. R. and Boulanger, F. and Bucher, M. and Burigana, C. and Butler, R. C. and Calabrese, E. and Cardoso, J.-F. and Catalano, A. and Challinor, A. and Chamballu, A. and Chary, R.-R. and Chiang, H. C. and Christensen, P. R. and Church, S. and Clements, D. L. and Colombi, S. and Colombo, L. P. L. and Combet, C. and Contreras, D. and Couchot, F. and Coulais, A. and Crill, B. P. and Curto, A. and Cuttaia, F. and Danese, L. and Davies, R. D. and Davis, R. J. and De Bernardis, P. and De Rosa, A. and De Zotti, G. and Delabrouille, J. and Désert, F.-X. and Diego, J. M. and Dole, H. and Donzelli, S. and Doré, O. and Douspis, M. and Ducout, A. and Dupac, X. and Efstathiou, G. and Elsner, F. and Enßlin, T. A. and Eriksen, H. K. and Fergusson, J. and Finelli⋆, F. and Forni, O. and Frailis, M. and Fraisse, A. A. and Franceschi, E. and Frejsel, A. and Frolov, A. and Galeotta, S. and Galli, S. and Ganga, K. and Gauthier, C. and Giard, M. and Giraud-Héraud, Y. and Gjerløw, E. and González-Nuevo, J. and Górski, K. M. and Gratton, S. and Gregorio, A. and Gruppuso, A. and Gudmundsson, J. E. and Hamann, J. and Handley, W. and Hansen, F. K. and Hanson, D. and Harrison, D. L. and Henrot-Versillé, S. and Hernández-Monteagudo, C. and Herranz, D. and Hildebrandt, S. R. and Hivon, E. and Hobson, M. and Holmes, W. A. and Hornstrup, A. and Hovest, W. and Huang, Z. and Huffenberger, K. M. and Hurier, G. and Jaffe, A. H. and Jaffe, T. R. and Jones, W. C. and Juvela, M. and Keihänen, E. and Keskitalo, R. and Kim, J. and Kisner, T. S. and Kneissl, R. and Knoche, J. and Kunz, M. and Kurki-Suonio, H. and Lagache, G. and Lähteenmäki, A. and Lamarre, J.-M. and Lasenby, A. and Lattanzi, M. and Lawrence, C. R. and Leonardi, R. and Lesgourgues, J. and Levrier, F. and Lewis, A. and Liguori, M. and Lilje, P. B. and Linden-Vørnle, M. and López-Caniego, M. and Lubin, P. M. and Ma, Y.-Z. and Macías-Pérez, J. F. and Maggio, G. and Maino, D. and Mandolesi, N. and Mangilli, A. and Maris, M. and Martin, P. G. and Martínez-González, E. and Masi, S. and Matarrese, S. and McGehee, P. and Meinhold, P. R. and Melchiorri, A. and Mendes, L. and Mennella, A. and Migliaccio, M. and Mitra, S. and Miville-Deschênes, M.-A. and Molinari, D. and Moneti, A. and Montier, L. and Morgante, G. and Mortlock, D. and Moss, A. and Münchmeyer, M. and Munshi, D. and Murphy, J. A. and Naselsky, P. and Nati, F. and Natoli, P. and Netterfield, C. B. and Nørgaard-Nielsen, H. U. and Noviello, F. and Novikov, D. and Novikov, I. and Oxborrow, C. A. and Paci, F. and Pagano, L. and Pajot, F. and Paladini, R. and Pandolfi, S. and Paoletti, D. and Pasian, F. and Patanchon, G. and Pearson, T. J. and Peiris, H. V. and Perdereau, O. and Perotto, L. and Perrotta, F. and Pettorino, V. and Piacentini, F. and Piat, M. and Pierpaoli, E. and Pietrobon, D. and Plaszczynski, S. and Pointecouteau, E. and Polenta, G. and Popa, L. and Pratt, G. W. and Prézeau, G. and Prunet, S. and Puget, J.-L. and Rachen, J. P. and Reach, W. T. and Rebolo, R. and Reinecke, M. and Remazeilles, M. and Renault, C. and Renzi, A. and Ristorcelli, I. and Rocha, G. and Rosset, C. and Rossetti, M. and Roudier, G. and Rowan-Robinson, M. and Rubiño-Martín, J. A. and Rusholme, B. and Sandri, M. and Santos, D. and Savelainen, M. and Savini, G. and Scott, D. and Seiffert, M. D. and Shellard, E. P. S. and Shiraishi, M. and Spencer, L. D. and Stolyarov, V. and Stompor, R. and Sudiwala, R. and Sunyaev, R. and Sutton, D. and Suur-Uski, A.-S. and Sygnet, J.-F. and Tauber, J. A. and Terenzi, L. and Toffolatti, L. and Tomasi, M. and Tristram, M. and Trombetti, T. and Tucci, M. and Tuovinen, J. and Valenziano, L. and Valiviita, J. and Van Tent, B. and Vielva, P. and Villa, F. and Wade, L. A. and Wandelt, B. D. and Wehus, I. K. and White, M. and Yvon, D. and Zacchei, A. and Zibin, J. P. and Zonca, A.},
	month = oct,
	year = {2016},
	pages = {A20},
}

@article{aghanim_planck_2020,
	title = {\textit{{Planck}} 2018 results: {VI}. {Cosmological} parameters},
	volume = {641},
	issn = {0004-6361, 1432-0746},
	shorttitle = {\textit{{Planck}} 2018 results},

	doi = {10.1051/0004-6361/201833910},

	journal = {Astronomy \& Astrophysics},
	author = {Aghanim, N. and Akrami, Y. and Ashdown, M. and Aumont, J. and Baccigalupi, C. and Ballardini, M. and Banday, A. J. and Barreiro, R. B. and Bartolo, N. and Basak, S. and Battye, R. and Benabed, K. and Bernard, J.-P. and Bersanelli, M. and Bielewicz, P. and Bock, J. J. and Bond, J. R. and Borrill, J. and Bouchet, F. R. and Boulanger, F. and Bucher, M. and Burigana, C. and Butler, R. C. and Calabrese, E. and Cardoso, J.-F. and Carron, J. and Challinor, A. and Chiang, H. C. and Chluba, J. and Colombo, L. P. L. and Combet, C. and Contreras, D. and Crill, B. P. and Cuttaia, F. and de Bernardis, P. and de Zotti, G. and Delabrouille, J. and Delouis, J.-M. and Di Valentino, E. and Diego, J. M. and Doré, O. and Douspis, M. and Ducout, A. and Dupac, X. and Dusini, S. and Efstathiou, G. and Elsner, F. and Enßlin, T. A. and Eriksen, H. K. and Fantaye, Y. and Farhang, M. and Fergusson, J. and Fernandez-Cobos, R. and Finelli, F. and Forastieri, F. and Frailis, M. and Fraisse, A. A. and Franceschi, E. and Frolov, A. and Galeotta, S. and Galli, S. and Ganga, K. and Génova-Santos, R. T. and Gerbino, M. and Ghosh, T. and González-Nuevo, J. and Górski, K. M. and Gratton, S. and Gruppuso, A. and Gudmundsson, J. E. and Hamann, J. and Handley, W. and Hansen, F. K. and Herranz, D. and Hildebrandt, S. R. and Hivon, E. and Huang, Z. and Jaffe, A. H. and Jones, W. C. and Karakci, A. and Keihänen, E. and Keskitalo, R. and Kiiveri, K. and Kim, J. and Kisner, T. S. and Knox, L. and Krachmalnicoff, N. and Kunz, M. and Kurki-Suonio, H. and Lagache, G. and Lamarre, J.-M. and Lasenby, A. and Lattanzi, M. and Lawrence, C. R. and Le Jeune, M. and Lemos, P. and Lesgourgues, J. and Levrier, F. and Lewis, A. and Liguori, M. and Lilje, P. B. and Lilley, M. and Lindholm, V. and López-Caniego, M. and Lubin, P. M. and Ma, Y.-Z. and Macías-Pérez, J. F. and Maggio, G. and Maino, D. and Mandolesi, N. and Mangilli, A. and Marcos-Caballero, A. and Maris, M. and Martin, P. G. and Martinelli, M. and Martínez-González, E. and Matarrese, S. and Mauri, N. and McEwen, J. D. and Meinhold, P. R. and Melchiorri, A. and Mennella, A. and Migliaccio, M. and Millea, M. and Mitra, S. and Miville-Deschênes, M.-A. and Molinari, D. and Montier, L. and Morgante, G. and Moss, A. and Natoli, P. and Nørgaard-Nielsen, H. U. and Pagano, L. and Paoletti, D. and Partridge, B. and Patanchon, G. and Peiris, H. V. and Perrotta, F. and Pettorino, V. and Piacentini, F. and Polastri, L. and Polenta, G. and Puget, J.-L. and Rachen, J. P. and Reinecke, M. and Remazeilles, M. and Renzi, A. and Rocha, G. and Rosset, C. and Roudier, G. and Rubiño-Martín, J. A. and Ruiz-Granados, B. and Salvati, L. and Sandri, M. and Savelainen, M. and Scott, D. and Shellard, E. P. S. and Sirignano, C. and Sirri, G. and Spencer, L. D. and Sunyaev, R. and Suur-Uski, A.-S. and Tauber, J. A. and Tavagnacco, D. and Tenti, M. and Toffolatti, L. and Tomasi, M. and Trombetti, T. and Valenziano, L. and Valiviita, J. and Van Tent, B. and Vibert, L. and Vielva, P. and Villa, F. and Vittorio, N. and Wandelt, B. D. and Wehus, I. K. and White, M. and White, S. D. M. and Zacchei, A. and Zonca, A.},
	month = sep,
	year = {2020},
	pages = {A6},
}

@article{akrami_planck_2020-1,
	title = {\textit{{Planck}} 2018 results: {X}. {Constraints} on inflation},
	volume = {641},
	issn = {0004-6361, 1432-0746},
	shorttitle = {\textit{{Planck}} 2018 results},

	doi = {10.1051/0004-6361/201833887},

	journal = {Astronomy \& Astrophysics},
	author = {Akrami, Y. and Arroja, F. and Ashdown, M. and Aumont, J. and Baccigalupi, C. and Ballardini, M. and Banday, A. J. and Barreiro, R. B. and Bartolo, N. and Basak, S. and Benabed, K. and Bernard, J.-P. and Bersanelli, M. and Bielewicz, P. and Bock, J. J. and Bond, J. R. and Borrill, J. and Bouchet, F. R. and Boulanger, F. and Bucher, M. and Burigana, C. and Butler, R. C. and Calabrese, E. and Cardoso, J.-F. and Carron, J. and Challinor, A. and Chiang, H. C. and Colombo, L. P. L. and Combet, C. and Contreras, D. and Crill, B. P. and Cuttaia, F. and de Bernardis, P. and de Zotti, G. and Delabrouille, J. and Delouis, J.-M. and Di Valentino, E. and Diego, J. M. and Donzelli, S. and Doré, O. and Douspis, M. and Ducout, A. and Dupac, X. and Dusini, S. and Efstathiou, G. and Elsner, F. and Enßlin, T. A. and Eriksen, H. K. and Fantaye, Y. and Fergusson, J. and Fernandez-Cobos, R. and Finelli, F. and Forastieri, F. and Frailis, M. and Franceschi, E. and Frolov, A. and Galeotta, S. and Galli, S. and Ganga, K. and Gauthier, C. and Génova-Santos, R. T. and Gerbino, M. and Ghosh, T. and González-Nuevo, J. and Górski, K. M. and Gratton, S. and Gruppuso, A. and Gudmundsson, J. E. and Hamann, J. and Handley, W. and Hansen, F. K. and Herranz, D. and Hivon, E. and Hooper, D. C. and Huang, Z. and Jaffe, A. H. and Jones, W. C. and Keihänen, E. and Keskitalo, R. and Kiiveri, K. and Kim, J. and Kisner, T. S. and Krachmalnicoff, N. and Kunz, M. and Kurki-Suonio, H. and Lagache, G. and Lamarre, J.-M. and Lasenby, A. and Lattanzi, M. and Lawrence, C. R. and Le Jeune, M. and Lesgourgues, J. and Levrier, F. and Lewis, A. and Liguori, M. and Lilje, P. B. and Lindholm, V. and López-Caniego, M. and Lubin, P. M. and Ma, Y.-Z. and Macías-Pérez, J. F. and Maggio, G. and Maino, D. and Mandolesi, N. and Mangilli, A. and Marcos-Caballero, A. and Maris, M. and Martin, P. G. and Martínez-González, E. and Matarrese, S. and Mauri, N. and McEwen, J. D. and Meerburg, P. D. and Meinhold, P. R. and Melchiorri, A. and Mennella, A. and Migliaccio, M. and Mitra, S. and Miville-Deschênes, M.-A. and Molinari, D. and Moneti, A. and Montier, L. and Morgante, G. and Moss, A. and Münchmeyer, M. and Natoli, P. and Nørgaard-Nielsen, H. U. and Pagano, L. and Paoletti, D. and Partridge, B. and Patanchon, G. and Peiris, H. V. and Perrotta, F. and Pettorino, V. and Piacentini, F. and Polastri, L. and Polenta, G. and Puget, J.-L. and Rachen, J. P. and Reinecke, M. and Remazeilles, M. and Renzi, A. and Rocha, G. and Rosset, C. and Roudier, G. and Rubiño-Martín, J. A. and Ruiz-Granados, B. and Salvati, L. and Sandri, M. and Savelainen, M. and Scott, D. and Shellard, E. P. S. and Shiraishi, M. and Sirignano, C. and Sirri, G. and Spencer, L. D. and Sunyaev, R. and Suur-Uski, A.-S. and Tauber, J. A. and Tavagnacco, D. and Tenti, M. and Toffolatti, L. and Tomasi, M. and Trombetti, T. and Valiviita, J. and Van Tent, B. and Vielva, P. and Villa, F. and Vittorio, N. and Wandelt, B. D. and Wehus, I. K. and White, S. D. M. and Zacchei, A. and Zibin, J. P. and Zonca, A.},
	month = sep,
	year = {2020},
	pages = {A10},
}

@article{racioppi_new_2018,
	title = {New universal attractor in nonminimally coupled gravity: {Linear} inflation},
	volume = {97},
	issn = {2470-0010, 2470-0029},
	shorttitle = {New universal attractor in nonminimally coupled gravity},

	doi = {10.1103/PhysRevD.97.123514},
	language = {en},
	number = {12},

	journal = {Physical Review D},
	author = {Racioppi, Antonio},
	month = jun,
	year = {2018},
	pages = {123514},
}

@article{riotto_inflation_2002,
	title = {Inflation and the {Theory} of {Cosmological} {Perturbations}},
	copyright = {arXiv.org perpetual, non-exclusive license},

	doi = {10.48550/ARXIV.HEP-PH/0210162},

	author = {Riotto, Antonio},
	year = {2002},
}

@article{senoguz_chaotic_2008,
	title = {Chaotic inflation, radiative corrections and precision cosmology},
	volume = {668},
	issn = {03702693},

	doi = {10.1016/j.physletb.2008.08.017},
	language = {en},
	number = {1},

	journal = {Physics Letters B},
	author = {Şenoğuz, V. Nefer and Shafi, Qaisar},
	month = sep,
	year = {2008},
	pages = {6--10},
}

@article{smoot_structure_1992,
	title = {Structure in the {COBE} {Differential} {Microwave} {Radiometer} {First}-{Year} {Maps}},
	volume = {396},
	issn = {0004-637X},

	doi = {10.1086/186504},

	journal = {The Astrophysical Journal},
	author = {Smoot, G. F. and Bennett, C. L. and Kogut, A. and Wright, E. L. and Aymon, J. and Boggess, N. W. and Cheng, E. S. and de Amici, G. and Gulkis, S. and Hauser, M. G. and Hinshaw, G. and Jackson, P. D. and Janssen, M. and Kaita, E. and Kelsall, T. and Keegstra, P. and Lineweaver, C. and Loewenstein, K. and Lubin, P. and Mather, J. and Meyer, S. S. and Moseley, S. H. and Murdock, T. and Rokke, L. and Silverberg, R. F. and Tenorio, L. and Weiss, R. and Wilkinson, D. T.},
	month = sep,
	year = {1992},
	note = {ADS Bibcode: 1992ApJ...396L...1S},
	pages = {L1},
}

@article{starobinsky_new_1980,
	title = {A new type of isotropic cosmological models without singularity},
	volume = {91},
	issn = {03702693},

	doi = {10.1016/0370-2693(80)90670-X},
	language = {en},
	number = {1},

	journal = {Physics Letters B},
	author = {Starobinsky, A.A.},
	month = mar,
	year = {1980},
	pages = {99--102},
}

@article{starobinsky_dynamics_1982,
	title = {Dynamics of phase transition in the new inflationary universe scenario and generation of perturbations},
	volume = {117},
	issn = {03702693},

	doi = {10.1016/0370-2693(82)90541-X},
	language = {en},
	number = {3-4},

	journal = {Physics Letters B},
	author = {Starobinsky, A.A.},
	month = nov,
	year = {1982},
	pages = {175--178},
}

@article{steigman_primordial_2007,
	title = {Primordial {Nucleosynthesis} in the {Precision} {Cosmology} {Era}},
	volume = {57},
	issn = {0163-8998, 1545-4134},

	doi = {10.1146/annurev.nucl.56.080805.140437},
	language = {en},
	number = {1},

	journal = {Annual Review of Nuclear and Particle Science},
	author = {Steigman, Gary},
	month = nov,
	year = {2007},
	pages = {463--491},
}

@article{traschen_particle_1990,
	title = {Particle production during out-of-equilibrium phase transitions},
	volume = {42},
	issn = {0556-2821},

	doi = {10.1103/PhysRevD.42.2491},
	language = {en},
	number = {8},

	journal = {Physical Review D},
	author = {Traschen, Jennie H. and Brandenberger, Robert H.},
	month = oct,
	year = {1990},
	pages = {2491--2504},
}

@article{turner_coherent_1983,
	title = {Coherent scalar-field oscillations in an expanding universe},
	volume = {28},
	issn = {0556-2821},

	doi = {10.1103/PhysRevD.28.1243},
	language = {en},
	number = {6},

	journal = {Physical Review D},
	author = {Turner, Michael S.},
	month = sep,
	year = {1983},
	pages = {1243--1247},
}

@article{Liddle_2003,
	doi = {10.1103/physrevd.68.103503},
  year = 2003,
	month = {nov},
  publisher = {American Physical Society ({APS})},
  volume = {68},
  number = {10},
  author = {Andrew R. Liddle and Samuel M. Leach},
  title = {How long before the end of inflation were observable perturbations produced?},
  
	journal = {Physical Review D},
}

@article{Dodelson_2003,
	doi = {10.1103/physrevlett.91.131301},
year = 2003,
	month = {sep},
  publisher = {American Physical Society ({APS})},
  volume = {91},
  number = {13},
  author = {Scott Dodelson and Lam Hui},
  title = {Horizon Ratio Bound for Inflationary Fluctuations},
  journal = {Physical Review Letters},
}

@article{Ghoshal:2022zwu,
	archiveprefix = "arXiv",
	author = "Ghoshal, Anish and Okada, Nobuchika and Paul, Arnab",
	doi = "10.1103/PhysRevD.106.095021",
	eprint = "2203.03670",
	journal = "Phys. Rev. D",
	number = "9",
	pages = "095021",
	primaryclass = "hep-ph",
	title = "{eV Hubble scale inflation with a radiative plateau: Very light inflaton, reheating, and dark matter in B-L extensions}",
	volume = "106",
	year = "2022"
}

@article{Drees:2021wgd,
	archiveprefix = "arXiv",
	author = "Drees, Manuel and Xu, Yong",
	doi = "10.1088/1475-7516/2021/09/012",
	eprint = "2104.03977",
	journal = "JCAP",
	pages = "012",
	primaryclass = "hep-ph",
	title = "{Small field polynomial inflation: reheating, radiative stability and lower bound}",
	volume = "09",
	year = "2021"
}

@article{Borah:2019bdi,
	archiveprefix = "arXiv",
	author = "Borah, Debasish and Nanda, Dibyendu and Saha, Abhijit Kumar",
	doi = "10.1103/PhysRevD.101.075006",
	eprint = "1904.04840",
	journal = "Phys. Rev. D",
	number = "7",
	pages = "075006",
	primaryclass = "hep-ph",
	title = "{Common origin of modified chaotic inflation, nonthermal dark matter, and Dirac neutrino mass}",
	volume = "101",
	year = "2020"
}

@article{Maity:2019ltu,
	archiveprefix = "arXiv",
	author = "Maity, Debaprasad and Saha, Pankaj",
	doi = "10.1088/1361-6382/ab0038",
	eprint = "1902.01895",
	journal = "Class. Quant. Grav.",
	pages = "045010",
	primaryclass = "gr-qc",
	title = "{Minimal plateau inflationary cosmologies and constraints from reheating}",
	volume = "36",
	year = "2019"
}

@article{Harada:2013epa,
    archivePrefix = "arXiv",
    author = "Harada, Tomohiro and Yoo, Chul-Moon and Kohri, Kazunori",
    doi = "10.1103/PhysRevD.88.084051",
    title = "{Threshold of primordial black hole formation}",
    eprint = "1309.4201",
   
    primaryClass = "astro-ph.CO",
    journal = "Phys. Rev. D",
    volume = "88",
    pages = "084051",
    year = "2013",
}
\end{document}